\documentclass[twocolumn,trackchanges]{aastex61hack}
\usepackage{natbib}
\usepackage{graphicx}

\newcommand\rp{\emph{r}-process}
\newcommand\rpe{\emph{r}-process-enhanced}
\newcommand\spr{\emph{s}-process}
\newcommand\rii{\emph{r}-II}
\newcommand\ri{\emph{r}-I}
\newcommand\limr{limited-\emph{r}}

\received{March 14, 2020}
\revised{June 4, 2020}
\accepted{June 6, 2020}

\shorttitle{RPA Fourth Data Release}
\shortauthors{Holmbeck et al.}

\begin{document}

\title{The \emph{R}-Process Alliance: Fourth Data Release from the Search for \emph{r}-Process-Enhanced Stars in the Galactic Halo\footnote{This paper includes data gathered with the 2.5 meter du Pont\\ telescope located at Las Campanas Observatory, Chile.}}

\correspondingauthor{Erika M.\ Holmbeck}
\email{eholmbec@nd.edu}

\author{Erika M.\ Holmbeck}
\affiliation{Department of Physics, University of Notre Dame, Notre Dame, IN 46556, USA}
\affiliation{Joint Institute for Nuclear Astrophysics -- Center for the Evolution of the Elements (JINA-CEE), USA}

\author{Terese T.\ Hansen}
\affiliation{George P.~and Cynthia Woods Mitchell Institute for Fundamental Physics and Astronomy, Texas A\&M University, College Station, TX 77843, USA}
\affiliation{Department of Physics and Astronomy, Texas A\&M University, College Station, TX 77843, USA}

\author{Timothy C.\ Beers}
\affiliation{Department of Physics, University of Notre Dame, Notre Dame, IN 46556, USA}
\affiliation{Joint Institute for Nuclear Astrophysics -- Center for the Evolution of the Elements (JINA-CEE), USA}

\author{Vinicius M.\ Placco}
\affiliation{Department of Physics, University of Notre Dame, Notre Dame, IN 46556, USA}
\affiliation{Joint Institute for Nuclear Astrophysics -- Center for the Evolution of the Elements (JINA-CEE), USA}

\author{Devin D.\ Whitten}
\affiliation{Department of Physics, University of Notre Dame, Notre Dame, IN 46556, USA}
\affiliation{Joint Institute for Nuclear Astrophysics -- Center for the Evolution of the Elements (JINA-CEE), USA}

\author{Kaitlin C.\ Rasmussen}
\affiliation{Department of Physics, University of Notre Dame, Notre Dame, IN 46556, USA}
\affiliation{Joint Institute for Nuclear Astrophysics -- Center for the Evolution of the Elements (JINA-CEE), USA}

\author{Ian U.\ Roederer}
\affiliation{Department of Astronomy, University of Michigan, Ann Arbor, MI 48109, USA}
\affiliation{Joint Institute for Nuclear Astrophysics -- Center for the Evolution of the Elements (JINA-CEE), USA}

\author{Rana Ezzeddine}
\affiliation{Department of Astronomy, University of Florida, Bryant Space Science Center, Gainesville, FL 32611, USA}
\affiliation{Joint Institute for Nuclear Astrophysics -- Center for the Evolution of the Elements (JINA-CEE), USA}

\author{Charli M.\ Sakari}
\affiliation{Department of Physics and Astronomy, San Francisco State University, San Francisco, CA 94132, USA}

\author{Anna Frebel}
\affiliation{Department of Physics and Kavli Institute for Astrophysics and Space Research, Massachusetts Institute of Technology, Cambridge, MA 02139, USA}
\affiliation{Joint Institute for Nuclear Astrophysics -- Center for the Evolution of the Elements (JINA-CEE), USA}

\author{Maria R.\ Drout}
\affiliation{Department of Astronomy and Astrophysics, University of Toronto, Toronto, ON, M5S 3H4, Canada}

\author{Joshua D.\ Simon}
\affiliation{The Observatories of the Carnegie Institution for Science, Pasadena, CA 91101, USA}

\author{Ian B.\ Thompson}
\affiliation{The Observatories of the Carnegie Institution for Science, Pasadena, CA 91101, USA}

\author{Joss Bland-Hawthorn}
\affiliation{Sydney Institute for Astronomy, School of Physics A28, University of Sydney, NSW 2006, Australia}

\author{Brad K.\ Gibson}
\affiliation{E.A.\ Milne Centre for Astrophysics, University of Hull, Hull, HU6 7RX, UK}

\author{Eva K.\ Grebel}
\affiliation{Astronomisches Rechen-Institut, Zentrum f\"ur Astronomie der Universit\"at Heidelberg, M\"onchhofstr.\ 12--14, 69120 Heidelberg, Germany}

\author{Georges Kordopatis}
\affiliation{Universit\'e C\^ote d'Azur, Observatoire de la C\^ote d'Azur, CNRS, Laboratoire Lagrange, Nice, France}

\author{Andrea Kunder}
\affiliation{Saint Martin's University, 5000 Abbey Way SE, Olympia, WA 98501, USA}

\author{Jorge Mel\'endez}
\affiliation{Instituto de Astronomia, Geof\'{i}sica e Ci\^{e}ncias Atmosf\'{e}ricas, Universidade de S\~{a}o Paulo, SP 05508-900, Brazil}

\author{Julio F.\ Navarro}
\affiliation{Department of Physics and Astronomy, University of Victoria, Victoria, BC, Canada V8P 5C2}

\author{Warren A.\ Reid}
\affiliation{Department of Physics and Astronomy, Macquarie University, Sydney NSW 2109, Australia}

\author{George Seabroke}
\affiliation{Mullard Space Science Laboratory, University College London, Holmbury St Mary, Dorking, RH5 6NT, UK}

\author{Matthias Steinmetz}
\affiliation{Leibniz Institut f\"ur Astrophysik Potsdam, An der Sterwarte 16, D-14482 Potsdam, Germany}

\author{Fred Watson}
\affiliation{Australian Government, Department of Industry, Innovation, Science, Energy and Resources, North Ryde, NSW 2113, Australia}

\author{Rosemary F.\. G.\ Wyse}
\affiliation{Johns Hopkins University, Dept of Physics \& Astronomy, Baltimore, MD 21218}

\begin{abstract}
This compilation is the fourth data release from the \emph{R}-Process Alliance (RPA) search for \rpe\ stars, and the second release based on ``snapshot" high-resolution ($R \sim 30,000$) spectra collected with the du Pont 2.5m Telescope.
In this data release, we propose a new delineation between the \ri\ and \rii\ stellar classes at $\mathrm{[Eu/Fe]} = +0.7$, instead of the empirically chosen $\mathrm{[Eu/Fe]} = +1.0$ level previously in use, based on statistical tests of the complete set of RPA data released to date. 
We also statistically justify the minimum level of [Eu/Fe] for definition of the \ri\ stars, [Eu/Fe] $> +0.3$.
Redefining the separation between \ri\ and \rii\ stars will aid in analysis of the possible progenitors of these two classes of stars and whether these signatures arise from separate astrophysical sources at all.
Applying this redefinition to previous RPA data, the number of identified \rii\ and \ri\ stars changes to 51 and 121, respectively, from the initial set of data releases published thus far.
In this data release, we identify 21 new \rii, 111 new \ri\ (plus three re-identified), and 7 new (plus one re-identified) \limr\ stars out of a total of 232 target stars, resulting in a total sample of 72 new \rii\ stars, 232 new \ri\ stars, and 42 new \limr\ stars identified by the RPA to date.
\end{abstract}


\section{Introduction}
\label{sec:intro}

Since the seminal work of \citet{burbidge1957} and \citet{cameron1957}, the rapid neutron-capture process (\rp) has been identified as a main physical mechanism responsible for the production of over half the elements in the Universe heavier than iron, with the other half produced primarily by the slow neutron-capture process (\spr).
Elemental production by the \spr\ likely occurs in shell burning in asymptotic giant branch (AGB) stars \citep{herwig2005,bisterzo2010}.
On the other hand, astrophysical sources that facilitate the production and release of \rp\ elements remain the topic of active research.
First proposed by \citet{lattimer1974}, neutron star mergers (NSMs) are currently favored as sites of the main \rp\ \citep{rosswog2014,thielemann2017}.
Observationally, NSMs gained support as sources of heavy \rp\ material with the inference of lanthanide material synthesized by an NSM associated with the gravitational wave signal detected by LIGO, GW170817 \citep{abbott2017,chornock2017,drout2017,kilpatrick2017,pian2017,shappee2017}.
It is still unclear whether NSMs are sufficiently frequent or prolific \rp\ sources to be responsible for the majority of \rp\ material in the Universe.

A prolific nucleosynthetic source occurring in the past leaves its elemental fingerprints on the Universe through stellar photospheres, which largely retain records of the gas from which the star formed.
In particular, very metal-poor (VMP; $[{\rm Fe/H}] < -2.0$) and extremely metal-poor (EMP; $[{\rm Fe/H}] < -3.0$) stars formed from gas that had not been enriched by many nucleosynthetic events prior to their birth.
A strong nucleosynthetic event enriching this chemically primitive metal-poor gas would leave a clear elemental signature in VMP and EMP stellar photospheres.
Indeed, at low metallicities (i.e., low [Fe/H]), distinct elemental signatures have been found over the past few decades, including stars enhanced with carbon---the so-called carbon-enhanced metal-poor, or CEMP stars \citep[see][]{beers2005}---and neutron-capture elements with a variety of patterns involving production by the  \spr, the \rp, and the recently suggested ``intermediate" (\emph{i}-) process \citep{cowan1977,dardelet2015,hampel2016,denissenkov2019}, the astrophysical site(s) of which are still under discussion.

Of particular importance are the \rpe\ stars, which exhibit enhancement of the heavy \rp\ elements ($Z\geq56$) in their photospheres.
The level of enrichment by the \rp\ in metal-poor stars is quantified by europium ($Z=63$), since this element is almost entirely produced by the \rp, and it is one of the easiest \rp\ elements to measure at optical wavelengths in stellar spectra.
Currently, the \rpe\ stars are divided into two sub-classes characterizing their enhancement: ``\ri," with $+0.3 < [{\rm Eu/Fe}] \leq +1.0$, and ``\rii," with $[{\rm Eu/Fe}] > +1.0$, corresponding to, respectively, a factor of over two and over ten times enriched compared to the Solar System \citep{beers2005}.
Among the VMP stars in the Galaxy, the \rii\ stars account for roughly 3--5\% and the \ri\ stars about 15--20\%, according to the limited amount of previously published work \citep{barklem2005}.
Recent dedicated survey efforts by the RPA find slightly higher rates of nearly 8\% of metal-poor stars displaying an \rii\ signature and 40\% an \ri\ \citep{hansen2018,roederer2018b,sakari2018a,sakari2018b,sakari2019,ezzeddine2020}.
The main \rp\ patterns of the \ri\ and \rii\ stars are nearly identical, differing only by a scaling factor.
It is currently unclear whether this difference in scaling is indicative of separate, more or less prolific \rp\ sources, or if the \ri\ and \rii\ stars share similar progenitors, but with the \ri\ stars suffering more dilution by the natal gas of their birth environments.

The \ri\ and \rii\ stars are believed to record clear elemental signatures of single---or a few---\rp\ events, offering a window into possible \rp\ sources, such as NSMs. The most metal-poor \ri\ and \rii\ stars were originally thought to be enriched by an \rp\ source occurring at very early times in the Galactic history, placing a timescale on \rp\ events.
Due to the short timescales required for the evolution of stars with masses $>$8--10~M$_\odot$, core-collapse supernovae (CCSNe) were originally thought to be natural \rp\ sources \citep{truran1978}, while the assumed long coalescence timescales for NSMs could not be accommodated with the expected short time required for the birth of the most metal-poor \rii\ stars (500 Myr to 1 Gyr).
One way in which NSMs have again gained support is through the discovery of the ultra-faint dwarf (UFD) galaxy Reticulum~II (Ret~II).
Of nine stars observed, \citet{ji2016a} and \citet{roederer2016} identified seven as \rii\ members---a much higher \rii\ fraction than that found in the general field populations of the Milky Way.
(Only high upper limits on [Eu/Fe] for the remaining two stars could be determined, which does not rule them out as additional \rii\ stars.)
The formation of \rpe\ stars in dwarf galaxy analogs of Ret~II may alleviate the tension with the metal-poor nature of the \rii\ stars and the long coalescence timescales of NSMs, depending on the rate of star formation in this galaxy.
A low-mass dwarf galaxy with few nucleosynthetic events will maintain its metal-poor nature longer than the Milky Way as a whole.
Furthermore, \citet{beniamini2016} and \citet{ji2016a} argue that the large number of CCSNe required could not simultaneously explain both the very low metallicity ($[\rm{Fe/H}] \sim -2.8$) and the strong \rp\ enrichment of Ret~II, and rather, indicate that a single high-yield event (e.g., an NSM) having occurred early in the star-formation history is more likely.
Another type of rare and high-yield event with \rp\ elements \citep[e.g., collapsars;][]{pruet2004,surman2004,siegel2018,miller2019} may also be responsible for the material in UFDs similar to Ret~II.
However, these alternative exotic sites have not yet been definitively observed to occur.

In addition, many studies support a hierarchical merger origin of the Milky Way halo stars \cite[e.g.,][and references therein]{freeman2002,bullock2005,zolotov2009,tumlinson2010,tissera2013}.
Given that limited amounts of dilution are required in order to maintain the large over-abundances of \rp\ elements following an \rp\ event in environments similar to UFDs like Ret II, it is natural that the \rpe\ metal-poor halo stars were also accreted from such small galaxies by the Milky Way.
\citet{roederer2018} investigated this hypothesis for highly \rpe\ stars in the halo by identifying dynamically linked groups of \rpe\ stars using data from the first RPA release \citep{hansen2018} and other sources.
These dynamical groups could have once been members of satellite galaxies that hosted a prolific \rp\ event prior to their disruption into the Galactic halo.
Additional explorations of this hypothesis are presently underway \citep[e.g.,][and Gudin et al., in prep.]{yuan2019}.

The abstract goal of the RPA is to understand the \rp, which cannot be done effectively with the handful of \rii\ stars that were identified before the RPA was established. Accordingly, Phase~II of the RPA is to identify 75--100 new \rii\ stars to build a robust catalog of observational \rp\ signatures with which to use in future analyses.
This Phase~II data release is an interim update on the RPA Search for \emph{R}-Process-Enhanced Stars in the Galactic Halo, expanding on the work of \citet{hansen2018}, \citet{sakari2018a}, and \citet{ezzeddine2020}.
In this phase, we obtain ``snapshot" (resolving power $R\sim25,000$--35,000 and signal-to-noise S/N $\sim$30) high-resolution spectra of stars that have been spectroscopically (or in some cases, photometrically) validated as metal-poor in previous studies with medium-resolution spectra \citep[see, e.g., RPA Phase~I][]{placco2018}.
This resolving power and S/N is sufficient for determining Sr, Ba, and Eu abundances (or meaningful upper limits) in order to identify and characterize the stars with \rp\ enrichment among our targets.
Using the previous definitions of the split between \ri\ and \rii\ stars, this data release adds four new \rii\ stars, 128 new \ri\ stars, and seven new limited-\emph{r} stars (of 232 total targets) to the cumulative progress of the RPA.
As we discuss below, it is now appropriate, based on the RPA data collected to date, to specify a different division point in [Eu/Fe] for the separation of \ri\ and \rii\ stars, thus revising these totals.


\section{Observations}
\label{observations}

The data in this fourth RPA data release---the third reporting snapshot, high-resolution spectroscopy taken with Southern Hemisphere telescopes---were obtained over a total of twenty nights in March, May, August, September, and November, 2017.

Target stars were selected after medium-resolution spectroscopic validation as metal-poor, and with effective temperatures useful for the identification of \rp\ elements (generally 4250 $< \rm{T_{eff}} < 5750$\,K), e.g., as reported by \citet{placco2018,placco2019}.
Prior to medium-resolution validation, most targets were originally selected using the criteria described in \citet{melendez2016} from the RAdial Velocity Experiment \citep[{\sc Rave};][]{steinmetz2006,kordopatis2013,matijevic2017,kunder2017}, and others were drawn from surveys such as SkyMapper \citep{wolf2018}, Best \& Brightest \citep[B\&B;][]{schlaufman2014}, Hamburg/ESO \citep{christlieb2008}, and the Large Sky Area Multi-Object Fibre Spectroscopic Telescope \citep[LAMOST;][]{deng2012}.

\begin{deluxetable*}{lrrrrrrrrl}
\tabletypesize{\scriptsize}
\tablecaption{Observation Log \label{tab:obslog}}
\tablehead{
\colhead{2MASS Stellar ID} & \colhead{RA} & \colhead{Dec} & \colhead{$V$ mag\tablenotemark{a}}
&\colhead{MJD} &\colhead{Exp} & \colhead{S/N}  & \colhead{RV$_{\rm helio}$}
& \colhead{RV$_{\rm err}$} &\colhead{Source\tablenotemark{b}}\\
\colhead{} & \colhead{} & \colhead{} & \colhead{}
&\colhead{} &\colhead{(s)} & \colhead{4129\,{\AA} } & \colhead{(km~s$^{-1}$)}
& \colhead{(km~s$^{-1}$)}&\colhead{}}
\startdata 
J00002416$-$1107454  &  00 00 24.0  &  $-$11 07 44.4  &  12.0  &  58080.06763  &  3123  &   40  &  $-$106.81  &  0.21  &  S \\
J00023429$-$1924590  &  00 02 34.3  &  $-$19 24 59.0  &  10.9  &  58077.04492  &  1100  &   28  &  $-$100.36  &  0.22  &  R \\
J00041581$-$5815524  &  00 04 15.8  &  $-$58 15 52.5  &  10.9  &  58075.05943  &  1100  &   25  &  $+$184.69  &  0.27  &  R \\
J00062986$-$5049319  &  00 06 29.8  &  $-$50 49 30.0  &  10.5  &  58074.10377  &  906  &   42  &  $+$214.56  &  0.39  &  SH \\
J00093394$-$1857008  &  00 09 34.0  &  $-$18 57 01.1  &  11.2  &  58081.06715  &  1200  &   46  &  $-$67.34   &  0.23  &  R \\
J00154806$-$6253207  &  00 15 48.1  &  $-$62 53 20.7  &  11.0  &  58075.02570  &  1200  &   28  &  $+$204.55  &  0.44  &  R \\
J00172430$-$3333151  &  00 17 24.3  &  $-$33 33 15.1  &  12.2  &  57985.00459  &  1200  &   25  &  $-$17.33   &  0.28  &  R  \\
J00182832$-$3900338  &  00 18 28.3  &  $-$39 00 32.4  &  11.2  &  58076.02986  &  1400  &   31  &  $+$346.12  &  0.21  &  R  \\
J00223225$-$4839449  &  00 22 32.2  &  $-$48 39 43.2  &  11.1  &  58075.04273  &  1200  &   27  &  $+$243.54  &  0.20  &  R \\
J00374325$-$1204391  &  00 37 43.3  &  $-$12 04 39.2  &  11.1  &  57985.00459  &  800  &   35  &  $-$27.84   &  0.41  &  R  \\
\enddata 
\tablenotetext{a}{{\sc RAVE} DR5 $V$ magnitudes are from \citet{munari2014}, B\&B are from \citet{henden2014}}
\tablenotetext{b}{R: {\sc RAVE} \citep{kordopatis2013,kunder2017}, B: B\&B \citep{schlaufman2014}, L: LAMOST \citep{deng2012}, S: SkyMapper \citep{wolf2018}, M: \citet{melendez2016}, H: Hamburg/ESO \citep{christlieb2008}, D: SAGA Database \citep{suda2017}.}
\tablenotetext{}{(This table is available in its entirety in machine-readable form.)}
\end{deluxetable*}

High-resolution ($R\sim30,000$) spectra were obtained with the Echelle spectrograph on the du Pont 2.5m telescope at the Las Campanas Observatory, using the 1{\arcsec}$\times$4{\arcsec} slit and 2$\times$1 on-chip binning.
The spectra cover a wavelength range from 3860\,{\AA} to 9000\,{\AA} for our 232 relatively bright stars ($10 \lesssim V \lesssim 13$) with low metallicities ($-3 \lesssim \mathrm{[Fe/H]} \lesssim -1$).
Data were reduced using the Carnegie Python Distribution\footnote{http://code.obs.carnegiescience.edu/} \citep[\texttt{CarPy};][]{kelson1998,kelson2000,kelson2003}.
Heliocentric radial velocities (RVs) were measured with the \emph{fxcor} task in the Image Reduction and Analysis Facility \citep[\texttt{IRAF}\footnote{\texttt{IRAF} is distributed by the National Optical Astronomy Observatory, which is operated by the Association of Universities for Research in Astronomy, Inc., under cooperative agreement with the NSF.}][]{tody1986,tody1993}, using order-by-order cross-correlation between the target and select RV standards: HD~14412 ($7.46$ km\,s$^{-1}$), HD~96700 ($12.84$ km\,s$^{-1}$), HD~146775 ($-30.15$ km\,s$^{-1}$), HD~22879 ($120.40$ km\,s$^{-1}$), and HD~189625 ($-28.13$ km\,s$^{-1}$), from \citet{soubiran2013}.
For each target, the RV is found by taking the weighted average of each order's individual radial-velocity measurements, following the iterative removal of 2-$\sigma$ outliers.
On average, 15 orders with strong, unsaturated features were used for cross-correlation of each spectrum.
The uncertainties on our measured RVs are calculated from the standard error of the mean of the individual order-by-order cross-correlation results from \textit{fxcor}.
The S/N per resolution element of each spectrum in the region of the 4129\,{\AA} \ion{Eu}{2} line was estimated by taking the square root of the total continuum counts.
A S/N of $\sim$30 at 4129\,{\AA} is sufficient for the Phase\,II snapshot determination of Eu abundances.
The computed heliocentric RVs and S/N for each target are listed in Table~\ref{tab:obslog}, along with the RA, DEC, $V$ magnitude, MJD of the observation, and the exposure times.
    
\section{Stellar Parameter Derivations and Abundance Analysis}
\label{analysis}

\subsection{Atmospheric Parameters}

\begin{deluxetable*}{lDrrrrr}
\tablecaption{\ion{Fe}{1} and \ion{Fe}{2} Equivalent Width Measurements\label{tab:ew_all}}
\tablehead{
\colhead{2MASS Stellar ID} & \multicolumn2c{Species} & \colhead{$\lambda$} & \colhead{$\chi$} & \colhead{$\log gf$} & \colhead{EW} & \colhead{$\log\epsilon$} \\
\colhead{} & \multicolumn2c{} & \colhead{(\AA)} & \colhead{(eV)} & \colhead{} & \colhead{(m\AA)} & \colhead{}}
\decimals
\startdata
J00002416$-$1107454 & 26.0 & 3948.10 & 3.24 & $-$0.59 & 62.0 & 5.06 \\
J00002416$-$1107454 & 26.0 & 3977.74 & 2.20 & $-$1.12 & 90.7 & 4.95 \\
J00002416$-$1107454 & 26.0 & 4001.66 & 2.17 & $-$1.90 & 64.6 & 5.13 \\
J00002416$-$1107454 & 26.0 & 4032.63 & 1.48 & $-$2.38 & 70.1 & 4.88 \\
J00002416$-$1107454 & 26.0 & 4058.22 & 3.21 & $-$1.18 & 29.7 & 5.01 \\
J00002416$-$1107454 & 26.0 & 4067.98 & 3.21 & $-$0.53 & 61.8 & 4.94 \\
\multicolumn{1}{c}{\vdots} & \multicolumn{2}{c}{\vdots} & \multicolumn{1}{c}{\vdots} & \multicolumn{1}{c}{\vdots} & \multicolumn{1}{c}{\vdots} & \multicolumn{1}{c}{\vdots} & \multicolumn{1}{c}{\vdots} \\
J00002416$-$1107454 & 26.1 & 5316.62 & 3.15 & $-$1.78 & 98.0 & 5.35 \\
J00002416$-$1107454 & 26.1 & 5325.55 & 3.22 & $-$3.16 & 19.1 & 5.36 \\
J00002416$-$1107454 & 26.1 & 5362.87 & 3.20 & $-$2.62 & 43.6 & 5.30 \\
J00002416$-$1107454 & 26.1 & 5534.85 & 3.25 & $-$2.87 & 21.7 & 5.17 \\
\enddata
\tablenotetext{}{(This table is available in its entirety in machine-readable form.)}
\end{deluxetable*}

For consistency between RPA data releases, we derive stellar parameters spectroscopically following RPA DR1 \citep{hansen2018}, in which the equivalent-width measurements of \ion{Fe}{1} and \ion{Fe}{2} lines are used to find the 1D LTE stellar parameters, based on ATLAS9 model atmospheres \citep{castelli2003}.
First, the equivalent widths (EWs) of a large number of Fe lines are measured (on average, 82 \ion{Fe}{1} and 20 \ion{Fe}{2} lines).
The \ion{Fe}{1} and \ion{Fe}{2} EWs are listed for each star in Table~\ref{tab:ew_all}.
Next, we use the 2017 version of \texttt{MOOG} \citep{sneden1973}, including the treatment of Rayleigh scattering described in  \citet{sobeck2011}\footnote{https://github.com/alexji/moog17scat}, to derive an Fe abundance for each line.
The effective temperature (${\rm T}_{\rm eff}$) is derived by minimizing the trend of \ion{Fe}{1} abundances as a function of transition excitation potential.
Spectroscopically derived atmospheric parameters using 1D LTE models systematically disagree with photometric determinations.
Therefore, to correct the offset between the spectroscopic and photometric temperature scales, we use the following relation from \citet{frebel2013} to adjust the temperature: 
    \[\mathrm{T}_{\rm eff, corrected} = \mathrm{T}_{\rm eff, initial} -0.1 \times \mathrm{T}_{\rm eff, initial} + 670.\]
As our sample is dominated by cool stars, this temperature shift is, on average,
about 200\,K, ranging from about 60\,K for the warmest stars and up to about 400\,K for the coolest stars.
The microturbulent velocity ($\xi$) is found by minimizing the abundance trend with reduced equivalent width.
Finally, the surface gravity ($\log g$) is adjusted until the average \ion{Fe}{2} abundance agrees with the \ion{Fe}{1} abundance, and the metallicity ([Fe/H]) is set by the \ion{Fe}{1} abundance.

    \begin{figure}[t]
    \includegraphics[width=\columnwidth]{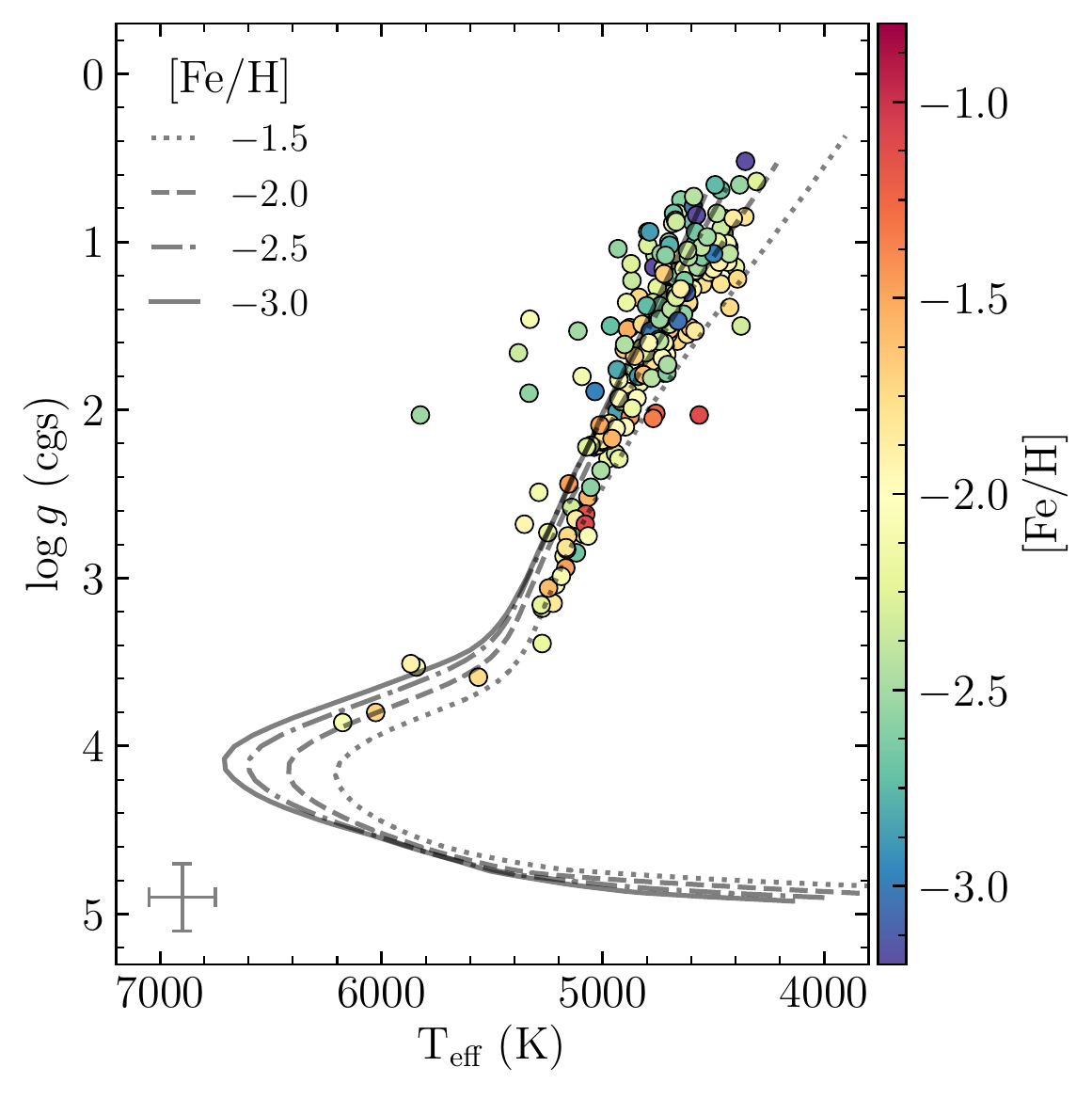}
    \caption{Surface gravity ($\log g$)  versus effective temperature  ($\rm{T}_{\rm eff}$) measurements for our
      target stars. Solid, dot-dashed, dashed, and dotted lines are 12-Gyr, $\alpha$-enhanced isochrones for $[{\rm Fe/H}]=-3.0$, $-2.5$, $-2.0$, and $-1.5$, respectively \citep{demarque2004}. The error bar in the lower-left corner represents standard uncertainties of 150~K and 0.2~dex on $\rm{T}_{\rm eff}$ and $\log g$, respectively.\label{fig:Tefflogg}} 
    \end{figure}

Assuming LTE can often underestimate the iron abundance relative to non-LTE, and therefore affect the determination of stellar parameters, especially for increasingly metal-poor stars.
Based on the low surface gravity and low temperature non-LTE \ion{Fe}{1} abundance correction models in \citet{lind2012}, we find that the average non-LTE correction to the [\ion{Fe}{1}/H] abundance is less than +0.2~dex for the stars in this sample that have $[{\rm Fe/H}]_{\rm LTE}$ between $-3$ and $-2$.
The correction decreases with both increasing metallicity and increasing surface gravity and becomes negligible at $[{\rm Fe/H}]>-1$.
This estimated correction is also supported by empirical fits to ultra-metal-poor stars in \citet{ezzeddine2017}, which show that for lower-metallicity stars, $[{\rm Fe/H}]_{\rm LTE}$ between $-3$ and $-2$, the non-LTE correction to the iron abundance can range between 0.13 and 0.27~dex, and a star with $[{\rm Fe/H}]\approx -1.0$ has a negligible $-0.01$~dex correction.
However, for lower-metallicity stars with $[{\rm Fe/H}]\approx -3.0$, this correction increases to nearly 0.3~dex, which dominates over the statistical line-by-line uncertainty.
As most of the targets in this sample have metallicities between $-3\leq [{\rm Fe/H}]_{\rm LTE}\leq -2$, the non-LTE effect to the iron abundance is at least on the order of the statistical line-by-line uncertainty and can significantly affect the derived atmospheric parameters.
To be consistent with previous RPA data releases, we assume LTE in the current study.
However, in future RPA data analyses, non-LTE effects will be homogeneously incorporated into the iron abundances and stellar parameter determinations.

\begin{deluxetable*}{l r r r r r r r}
\tablecaption{Model Atmospheric Parameters \label{tab:params}}
\tablewidth{0pt}
\tablehead{
\colhead{2MASS Stellar ID} & \colhead{T$_{\rm eff}$} & \colhead{$\log g$}
& \colhead{[Fe/H]} & \colhead{$\sigma_{\rm [Fe/H]}$}
& \colhead{$N$ \ion{Fe}{1}} & \colhead{$N$ \ion{Fe}{2}}
& \colhead{$\xi$} \\
\colhead{} & \colhead{(K)} & \colhead{(cgs)}
& \colhead{} & \colhead{}
& \colhead{} & \colhead{}
& \colhead{(km~s$^{-1}$)}}
\startdata 
J00002416$-$1107454  & 4693 & 1.37 & $-$2.40 & 0.12 &  90 &  26 & 2.11 \\ 
J00023429$-$1924590  & 4400 & 1.15 & $-$2.22 & 0.14 &  32 &  15 & 2.97 \\ 
J00041581$-$5815524  & 4375 & 1.50 & $-$2.32 & 0.17 &  30 &  12 & 2.81 \\ 
J00062986$-$5049319  & 4647 & 0.75 & $-$2.59 & 0.15 &  86 &  27 & 2.36 \\ 
J00093394$-$1857008  & 4815 & 1.78 & $-$1.85 & 0.14 & 115 &  27 & 1.56 \\ 
J00154806$-$6253207  & 4725 & 1.78 & $-$2.30 & 0.15 &  60 &  16 & 2.09 \\ 
J00172430$-$3333151  & 4764 & 1.73 & $-$2.29 & 0.13 &  69 &  16 & 2.02 \\ 
J00182832$-$3900338  & 4639 & 1.34 & $-$1.75 & 0.13 &  60 &  17 & 2.09 \\ 
J00223225$-$4839449  & 4648 & 1.40 & $-$1.75 & 0.15 & 102 &  21 & 2.15 \\
J00374325$-$1204391  & 4695 & 1.31 & $-$2.40 & 0.13 &  98 &  23 & 1.98 \\ 
\enddata
\tablenotetext{}{(This table is available in its entirety in machine-readable form.)}
\end{deluxetable*}

The (LTE) model atmospheric parameters are listed in Table \ref{tab:params}.
Figure \ref{fig:Tefflogg} shows the final derived surface gravities as a function of the derived effective temperature (after the applied correction) compared to 12~Gyr, $\alpha$-enhanced, metal-poor isochrones for a 0.8~M$_\odot$ star at different metallicities \citep{demarque2004}, showing that our sample is mainly comprised of metal-poor giants and validating our medium-resolution efforts.
These isochrones do not extend to the hot and low-gravity AGB region, where some of our target stars populate Figure~\ref{fig:Tefflogg}.
A few of our target stars were more metal-rich than previously estimated from the medium-resolution spectroscopic validation described in \citet{placco2018}, but overall that method was effective for identifying metal-poor stars.

\subsection{Abundances}

    \begin{figure*}
    \center
    \includegraphics[width=0.85\textwidth]{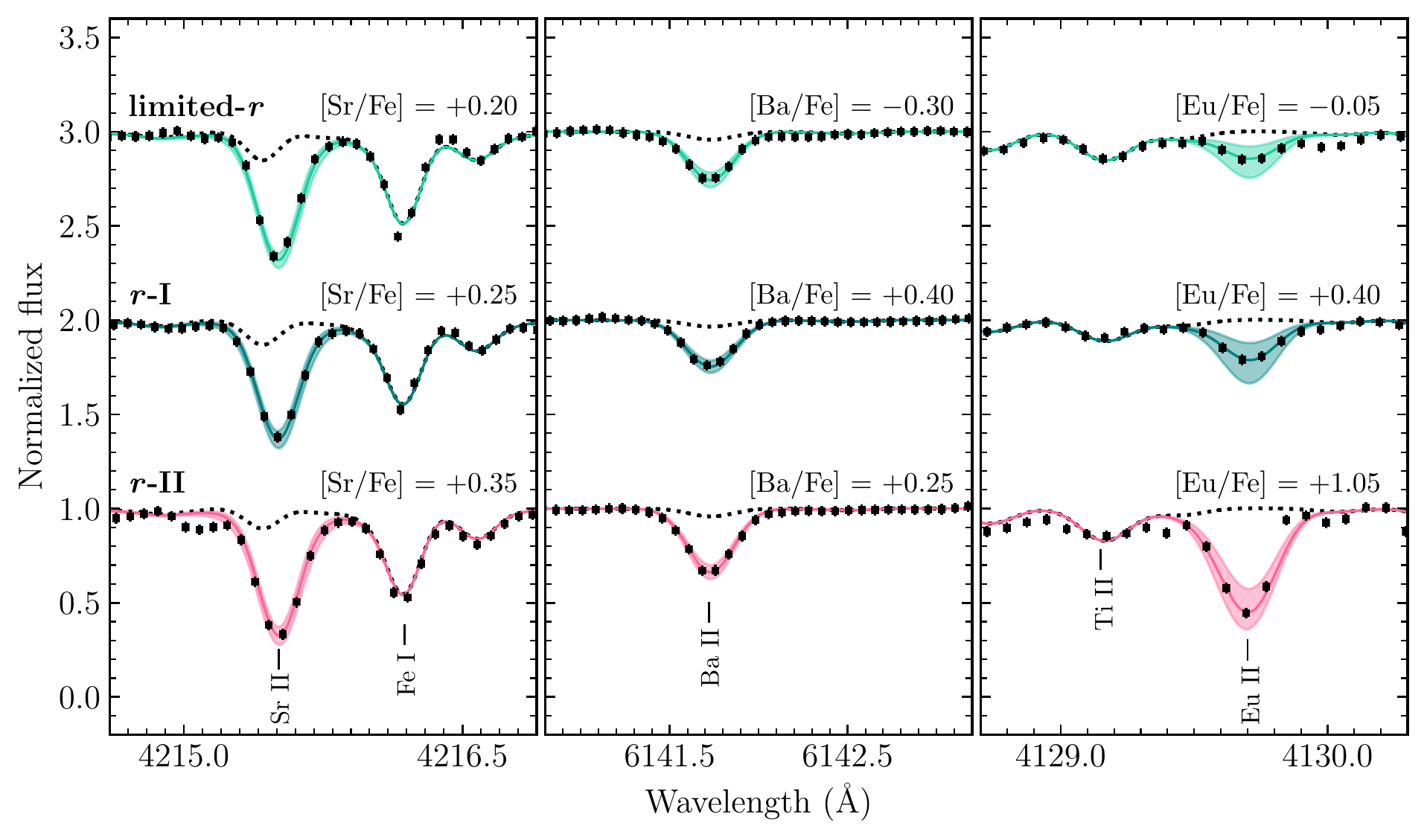}
    \caption{Scaled spectrum snippets (points) in the regions of interest for a \limr\ star (top, J10344785$-$4823544), an \ri\ star (middle, J20194310$-$3158163), and an \rii\ star (bottom, J03422816$-$6500355). The \texttt{MOOG} syntheses for \ion{Sr}{2} (left), \ion{Ba}{2} (middle), and \ion{Eu}{2} (right) are shown for no abundance (dashed line) and the best-fit case (colored line), with a conservative $\pm$0.30~dex uncertainty (shaded), which more than accommodates the random uncertainty due to S/N and systematic uncertainties in the atmospheric parameters.\label{fig:spec}}
    \end{figure*}

We derive abundances for C, Sr, Ba, and Eu from spectral synthesis using \texttt{MOOG}, in order to make an initial classification of each target into either \ri, \rii, \limr, CEMP, or no \rp\ enhancement (``non-RPE").
For the estimation of the stellar abundances, we use $\alpha$-enhanced ($[\alpha/\textrm{Fe}]= +0.4$) ATLAS9 model atmospheres \citep{castelli2003}.  
Line lists for each region of interest are generated with \texttt{linemake}\protect\footnote{\url{https://github.com/vmplacco/linemake}}.
These line lists include CH, C$_2$, and CN  molecular lines \citep{brooke2013,masseron2014,ram2014,sneden2014}, as well as isotopic shift and hyperfine-structure information for Ba and Eu \citep{lawler2001,gallagher2010}.
We use the Solar isotopic ratios in \citet{sneden2008} for neutron-capture elements with hyperfine-splitting effects.

The C abundances were primarily derived by fitting the entire CH $G$-band at 4313\,{\AA}. 
For cooler CEMP stars, where the $G$-band is saturated, abundances were derived from the C$_{2}$ Swan band at 5161\,{\AA}.
The Sr abundances were derived from two strong lines, at $\lambda$4077\,{\AA} and $\lambda$4215\,{\AA}, which can be significantly blended with Fe (and $_{66}$Dy, if present).
We derive Ba abundances from lines at $\lambda$5853\,{\AA}, $\lambda$6141\,{\AA}, and $\lambda$6496\,{\AA}.
Eu abundances are mainly derived from the $\lambda$4129\,{\AA}, $\lambda$4205\,{\AA}, and $\lambda$4435\,{\AA} features.
Since the $\lambda$4435\,{\AA} line is heavily blended by a neighboring Fe feature, and $\lambda$4205\,{\AA} by C and Ca, the $\lambda$4129\,{\AA} feature is primarily used to derive the Eu abundance.
The $\lambda$4205\,{\AA} feature may be significantly blended with C if the target is C-enhanced.
However, since most of our targets do not have enhanced C, the $\lambda$4205\,{\AA} line was minimally affected.
Figure~\ref{fig:spec} shows key Sr, Ba, and Eu features in a \limr, \ri, and \rii\ star along with their synthesized abundance.

\subsection{Abundance Uncertainties}

In this section, we estimate the uncertainties on the derived abundances from constant stellar parameter uncertainties.
First, we assume a conservative typical uncertainty on effective temperature of 150~K, 0.2~dex on surface gravity, and 0.2~km\,s$^{-1}$ on microturbulence.
We do not vary the metallicity, but instead use the random uncertainty associated with the line-by-line variation between iron abundances, i.e., $\sigma_{\rm [Fe/H]}$ in Table~\ref{tab:params}.
Then, we choose spectra that represent the parameter ranges for our targets, i.e., a somewhat hot star ($\sim$5000~K) with $[\rm{Fe/H}]\approx-2.0$, a cooler star ($\sim$4500~K) star with $[\rm{Fe/H}]\approx-2.5$, and a moderate-temperature ($\sim$4800~K) with $\log g\approx 1.0$.
With these three representative targets, we vary each of the stellar parameters within the above uncertainties individually and rederive the best-fit abundances for C, Sr, Ba, and Eu.

\begin{deluxetable*}{lrrrrrr}
\tablecaption{Abundances Uncertainties for Example Stars\label{tab:errors}}
\tablehead{
\colhead{} & \colhead{T$_{\rm eff}$} & \colhead{$\log g$} & \colhead{$\xi$} & \colhead{[Fe/H]} & \colhead{$\sigma_{\rm sys}$} & \colhead{$\sigma_{\rm tot}$} \\
\colhead{} & \colhead{$\pm$150~K} & \colhead{$\pm$0.2~dex} & \colhead{$\pm$0.2~km\,s$^{-1}$} & \colhead{$\pm\sigma_{\rm [Fe/H]}$} & \colhead{} & \colhead{}}
\startdata
J16285613$-$1014576 & 5078 & 1.80 & 2.07 & $-$1.93 & & \\
\hline
{[C/Fe]}  & $\pm$0.23 & $\mp$0.05 & $\mp$0.05 & $\mp$0.11 & $\pm$0.24 & $\pm$0.26 \\
{[Sr/Fe]} & $\pm$0.11 & $\pm$0.06 & $\mp$0.04 & $\mp$0.11 & $\pm$0.13 & $\pm$0.17 \\
{[Ba/Fe]} & $\pm$0.11 & $\pm$0.07 & $\mp$0.12 & $\mp$0.11 & $\pm$0.18 & $\pm$0.21 \\
{[Eu/Fe]} & $\pm$0.07 & $\pm$0.06 & $\mp$0.03 & $\mp$0.11 & $\pm$0.10 & $\pm$0.15 \\
\hline
J20504869$-$3355289 & 4549 & 1.09 & 2.33 & $-$2.63 & & \\
\hline
{[C/Fe]}  & $\pm$0.32 & $\mp$0.10 & $\mp$0.03 & $\mp$0.14 & $\pm$0.34 & $\pm$0.36 \\
{[Sr/Fe]} & $\pm$0.15 & $\pm$0.04 & $\mp$0.11 & $\mp$0.14 & $\pm$0.19 & $\pm$0.24 \\
{[Ba/Fe]} & $\pm$0.08 & $\pm$0.05 & $\mp$0.08 & $\mp$0.14 & $\pm$0.12 & $\pm$0.19 \\
{[Eu/Fe]} & $\pm$0.10 & $\pm$0.05 & $\mp$0.01 & $\mp$0.14 & $\pm$0.11 & $\pm$0.18 \\
\hline
J04014897$-$3757533 & 4797 & 1.02 & 2.32 & $-$2.28 & & \\
\hline
{[C/Fe]}  & $\pm$0.33 & $\mp$0.08 & $\mp$0.05 & $\mp$0.13 & $\pm$0.34 & $\pm$0.37 \\
{[Sr/Fe]} & $\pm$0.10 & $\pm$0.06 & $\mp$0.14 & $\mp$0.13 & $\pm$0.18 & $\pm$0.22 \\
{[Ba/Fe]} & $\pm$0.08 & $\pm$0.07 & $\mp$0.08 & $\mp$0.13 & $\pm$0.13 & $\pm$0.19 \\
{[Eu/Fe]} & $\pm$0.08 & $\pm$0.05 & $\mp$0.02 & $\mp$0.13 & $\pm$0.10 & $\pm$0.16 \\
\enddata
\end{deluxetable*}

\begin{figure}[!t]
    \center
    \includegraphics[width=\columnwidth]{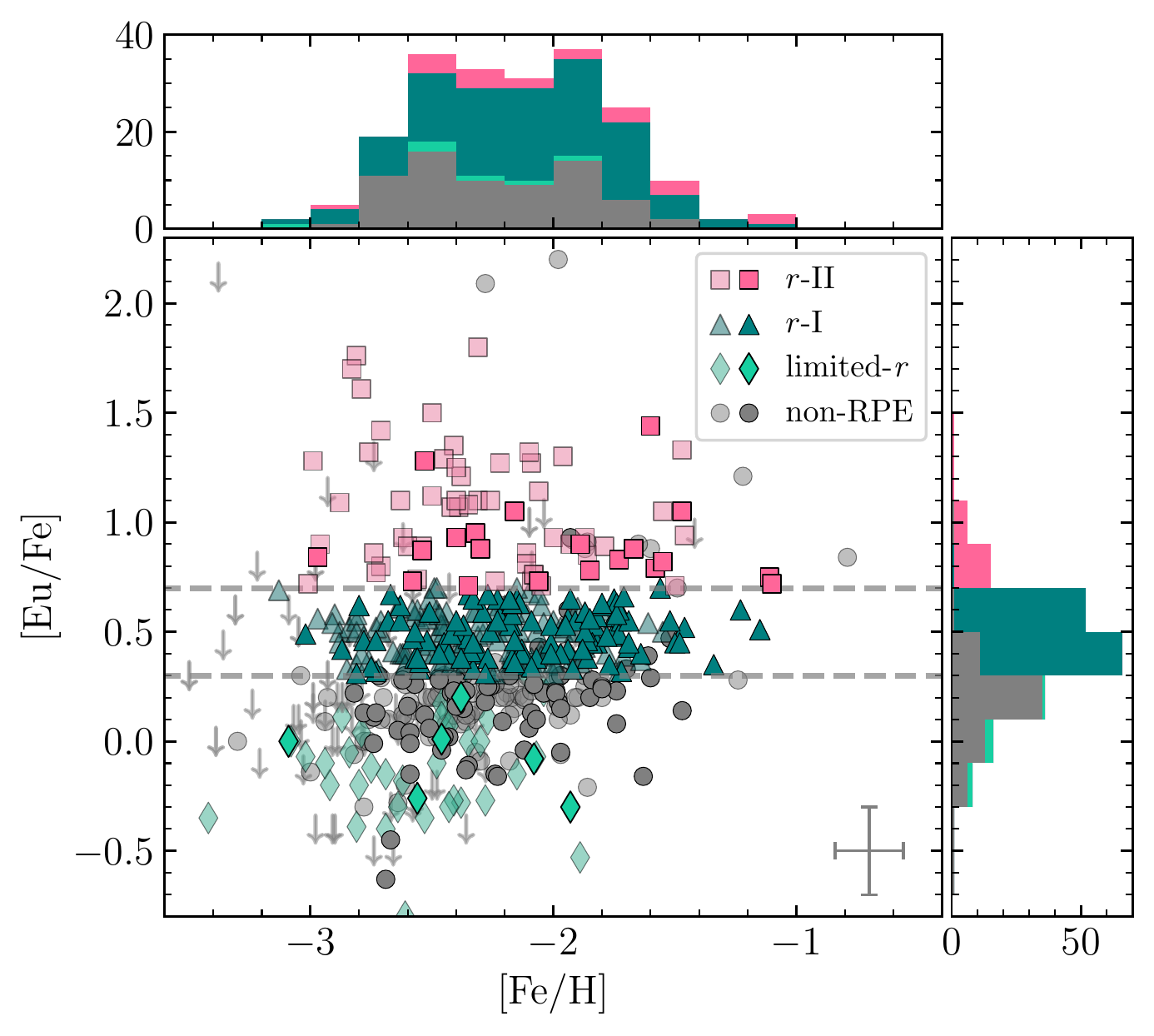}
    \caption{Derived $\mathrm{[Eu/Fe]}$ abundances as a function of metallicity for the stars in this sample, labeled by \rp\ enhancement type: non-RPE (circles), \rii\ (squares), \ri\ (triangles), and \limr\ (diamonds).
    Upper limits on [Eu/Fe] are indicated by a downward arrow.
    Also shown are the current RPA-identified \rp-enhanced stars (same labeling, lighter colors).
    Horizontal dashed lines indicate $\mathrm{[Eu/Fe]} = +0.3$ and $+0.7$, showing the new suggested cutoff levels on [Eu/Fe] for \ri\ and \rii\ classification. See \ref{sec:cutoff} for details.\label{fig:EuFe}} 
    \end{figure}

Table~\ref{tab:errors} reports the abundance variations after changing the atmospheric parameters individually.
We report both the systematic uncertainty ($\sigma_{\rm sys}$) from the atmospheric parameters as well as the total uncertainty when the random error on the metallicity is included ($\sigma_{\rm tot}$).
Note that it is more appropriate to use $\sigma_{\rm sys}$ when using the $\log\epsilon$ abundances and $\sigma_{\rm tot}$ for [X/Fe] abundances.
On average, the uncertainty on the [Sr/Fe], [Ba/Fe], and [Eu/Fe] abundances round to 0.2~dex.
The average random uncertainty from [Fe/H] is 0.14~dex for stars in this data release.
These average uncertainties are represented in Figures~\ref{fig:EuFe} and \ref{fig:SrBa} by an error bar in the corner.

\section{Results}
\label{results}

\begin{deluxetable*}{l r r r r r r l}
\tablecaption{Neutron-Capture Abundances and Sub-Class Assignments\label{tab:abund}}
\tablewidth{0pt}
\tablehead{
\colhead{2MASS Stellar ID} & \colhead{[Fe/H]} & \colhead{[C/Fe]} & \colhead{[C/Fe]$_{\rm c}$}
& \colhead{[Sr/Fe]} & \colhead{[Ba/Fe]} & \colhead{[Eu/Fe]}
& \colhead{Sub-class}}
\startdata
J00002416$-$1107454  &  $-$2.40  &  $-$0.33  &  $+$0.24  &  $-$0.32  &  $-$0.27  &  $+$0.50  &  \ri \\
J00023429$-$1924590  &  $-$2.22  &  $-$0.64  &  $+$0.06  &  $+$0.08  &  $-$0.27  &  $+$0.56  &  \ri \\
J00041581$-$5815524  &  $-$2.32  &  $-$0.66  &  $-$0.16  &  $+$0.72  &  $+$0.30  &  $+$0.95  &  \rii \\
J00062986$-$5049319  &  $-$2.59  &  $-$0.65  &  $+$0.11  &  $-$0.45  &  $-$0.70  &  $-$0.15  &  non-RPE \\
J00093394$-$1857008  &  $-$1.85  &  $-$0.17  &  $+$0.06  &  $+$0.13  &  $+$0.23  &  $+$0.46  &  \ri \\
J00154806$-$6253207  &  $-$2.30  &  $-$0.55  &  $-$0.33  &  $+$0.30  &  $+$0.08  &  $+$0.40  &  \ri \\
J00172430$-$3333151  &  $-$2.29  &  $-$0.07  &  $+$0.23  &  $+$0.35  &  $+$0.05  &  $+$0.59  &  \ri \\
J00182832$-$3900338  &  $-$1.75  &  $-$0.35  &  $+$0.14  &  $+$0.28  &  $+$0.07  &  $+$0.57  &  \ri \\
J00223225$-$4839449  &  $-$1.75  &  $-$0.25  &  $+$0.20  &  $-$0.05  &  $+$0.10  &  $+$0.65  &  \ri \\
J00374325$-$1204391  &  $-$2.40  &  $-$0.20  &  $+$0.42  &     0.00  &  $-$0.27  &  $+$0.28  &  non-RPE \\
\enddata
\tablenotetext{a}{Also analyzed in \citet{sakari2018a}.}
\tablenotetext{b}{\citet{casey2015} have also analyzed this star and find $[\rm{Eu/Fe}] < +0.50$.}
\tablenotetext{}{(This table is available in its entirety in machine-readable form.)}
\end{deluxetable*}

Final derived Fe, C, Sr, Ba, and Eu abundances for our 232 program stars are listed in Table \ref{tab:abund}, along with their \rp\ classification.
All [X/Y] abundances use the Solar System measurements from \citet{asplund2009}.
The neutron-capture elements---Sr, Ba, and Eu---indicate which neutron-capture processes dominated the elemental production preceding the formation of these stars; the absorption features of these elements are among the strongest of all neutron-capture elements for stars with similar atmospheric parameters.
Together, these five elements provide a comprehensive overview of the nucleosynthetic history of each star.
This work focuses especially on characterizing the neutron-capture nucleosynthesis signatures in halo stars.
    
The Sr, Ba, and Eu abundances are used to both determine the dominant source of the neutron capture elements and also to quantify various regions of the \rp\ pattern.
In particular, Ba and Eu abundances can be used as a metric for whether the neutron-capture elements in each star were primarily synthesized through an \emph{s-} or \rp\ \citep[][and references therein]{frebel2018}.
In essence, if the \rp\ dominated the production of neutron-capture elements, the observed ratio of Ba to Eu will be less than the Solar value, or $[\rm{Ba/Eu}] < 0$.
Alternatively, the $[\rm{Ba/Eu}]$ abundance is greater than Solar if the \spr\ dominated the production of neutron-capture elements.
In the neutron-capture elemental abundance pattern, the ``first \rp\ peak" is approximately indicated by the Sr abundance, while Ba is approximately representative of the second abundance peak.
Current studies argue that the light \rp\ elements (i.e., the first \rp\ peak) could originate from a separate \rp\ source (the limited \rp) than that which synthesized the second and third \rp\ peaks \citep{truran2002,honda2006,wanajo2006}.
Thus, the ratio of Sr to Ba quantifies the amount of \limr\ production.
These considerations motivate using the relative abundances of Sr, Ba, and Eu to determine whether the elements in each originated primarily from a \limr, \spr, or \rp\ production site.

\subsection{Revisiting the [Eu/Fe] \ri\ and \rii\ Cutoff Value}
\label{sec:cutoff}
    
Figure~\ref{fig:EuFe} shows the [Eu/Fe] abundances as a function of metallicity from this work and previous RPA data releases.
The majority of the targets were identified as \ri\ stars, with eight as \limr\ stars, under the current \ri\ and \limr\ definitions.
Using the value [Eu/Fe] $> +1.0$, as employed by the RPA up to now, only four new \rii\ stars were identified in the present data release.
The four stars with $\mathrm{[Eu/Fe]}>+1.0$ and $\mathrm{[Ba/Eu]} < 0.0$ are J03422816--6500355, J05383296--5904280, J07103110--7121522, and J07202253--3358518.
They are all of moderate temperature and together span about 1~dex in metallicity.
This rate ($\sim$1.7\%) indicates a decrease relative to the previous success rate of RPA efforts, which have either agreed with or exceeded the expected \rii\ discovery rate among VMP stars of 3--5\%, as estimated by \citet{christlieb2004} and \citet{barklem2005}.
This decrease is likely the result of the extension to higher metallicity of our present sample compared with previous RPA data releases.
    
The distribution of $\mathrm{[Eu/Fe]}$ abundances found in metal-poor stars is likely to be a continuum, unless different classes of \rp\ progenitors contribute significantly different amounts of lanthanides, which remains uncertain at present.
A simple Kolmogorov-Smirnov test fails to rule out the null hypothesis that \ri\ and \rii\ stars from the full RPA sample to date are drawn from the same parent distribution of [Fe/H], as has been previously speculated based on smaller samples \citep[e.g.,][]{barklem2005}.
Still, it is operationally useful to differentiate between moderately and extremely \rpe\ stellar signatures to investigate whether these stars have different \rp\ sources.
With the availability of the now myriad amount of data from RPA efforts, we can reconsider where this split between \ri\ and \rii\ stars should lie, based on the data in hand.

    \begin{figure}[t]
    \centering
    \includegraphics[width=0.9\columnwidth,trim={0 0.3cm 0.5cm 1cm},clip]{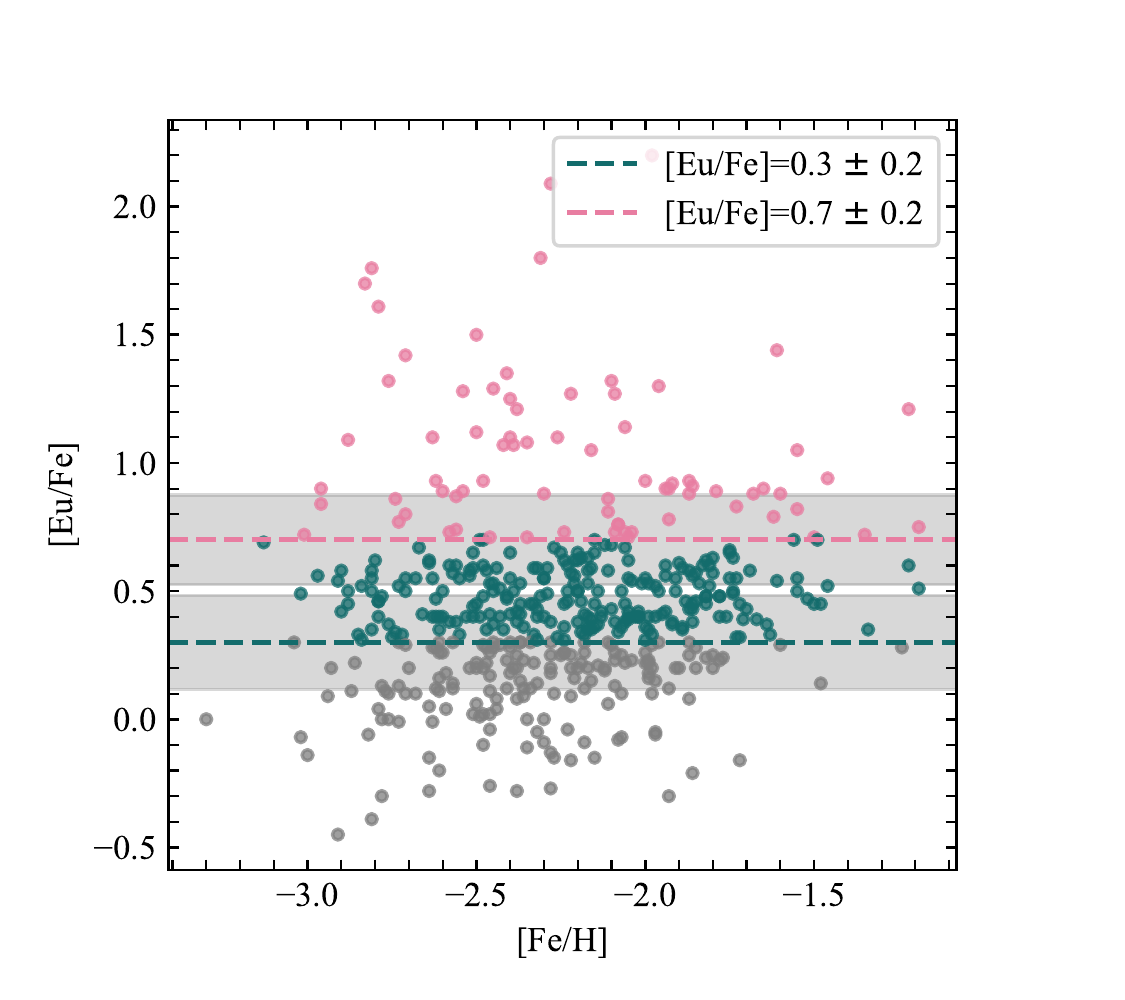}
    \includegraphics[width=0.9\columnwidth,trim={0 0.3cm 0.5cm 1cm},clip]{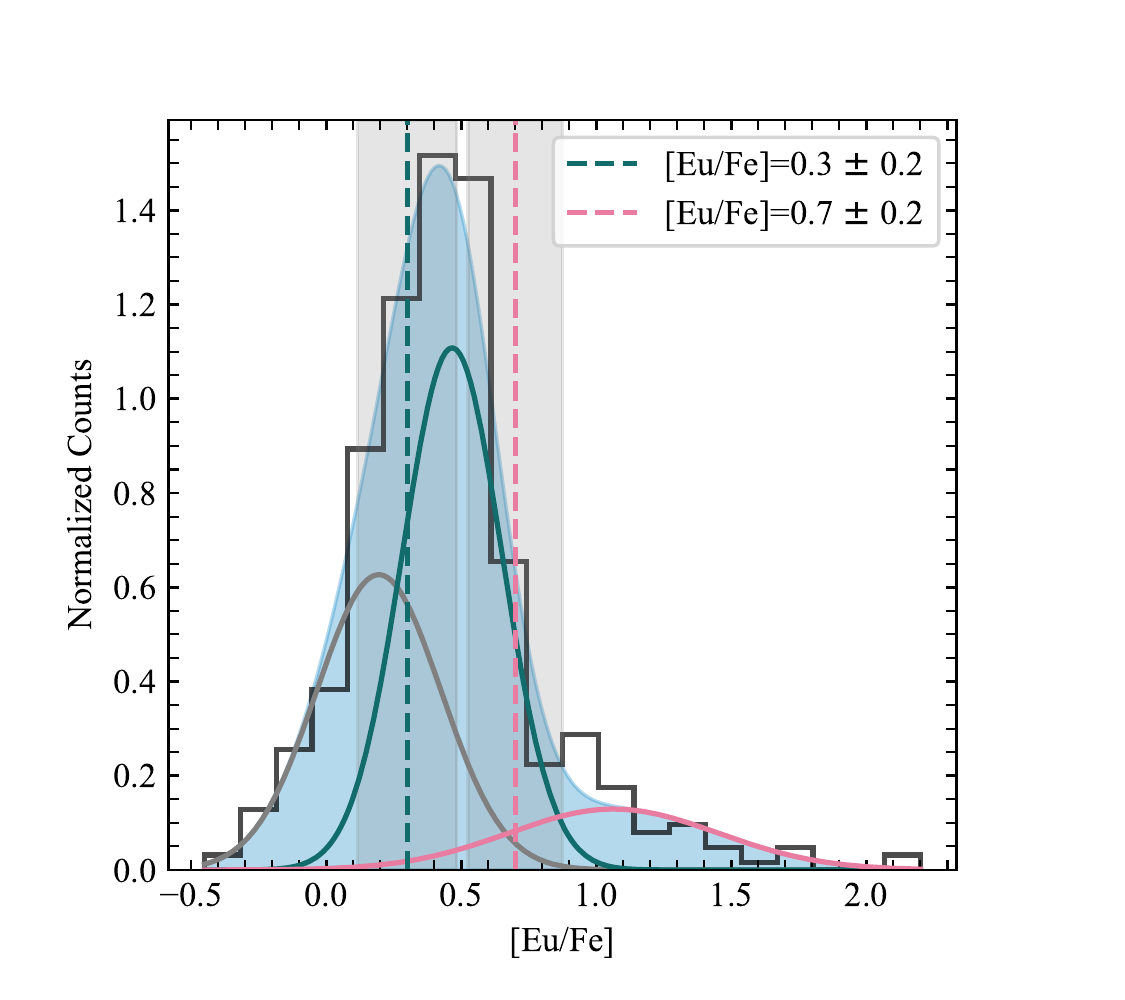}
    \caption{Top: scatter plot of [Eu/Fe] as a function of metallicity for RPA stars (excluding upper limits) colored by the average $k$-medoids grouping with $k$=3. Bottom: histogram of all RPA [Eu/Fe] abundances compared to Gaussian mixture model with three components. Teal and pink lines with gray shaded regions correspond to the average with their standard deviations of the $k$-medoids decision boundaries defining \ri\ and \rii.\label{fig:km_clustering}}
    \end{figure}

Without appeal to any particular physical models, we agnostically consider the existence of two or three distinct populations within the entire [Eu/Fe] distribution (note that we include the \limr\ stars for this exercise).
To mitigate concerns of the sample size ($N=471$) contributing to misinterpretation, we consider the \ri\ and \rii\ boundaries resulting from the robust partitioning technique known as $k$-medoids \citep{kaufman1990}.
Similar to the $k$-means algorithm, this partitioning procedure seeks to minimize the distance between cluster members to determine cluster centers.
We consider the cases $k$=2 and $k$=3 clusters separately, and evaluate the resulting [Eu/Fe] classifications.
In the case of two clusters, the boundary is determined to occur at $[\textrm{Eu/Fe}] = +0.4\pm0.2$, whereas the three-cluster case results in the boundaries $[\textrm{Eu/Fe}] = +0.3\pm0.1$ and $+0.7 \pm0.2$ for \ri\ and \rii\ classification, respectively.
Increasing the number of clusters did not significantly reduce the information loss, so we do not consider $k>3$ cases.

We evaluate the extent to which the [Eu/Fe] distribution is better represented by two or three components with a Gaussian mixture model via the Akaike information criterion \citep[AIC;][]{akaike1973}.
This criterion appropriately weights the goodness-of-fit with the simplicity of the model, mitigating the effects of overfitting when arbitrarily adding additional components to the model. 
Using a two-component Gaussian mixture model suggests a slightly higher degree of information loss ($\rm{AIC}=284$) than a three-component mixture ($\rm{AIC}=278$), from which we conclude that that sample [Eu/Fe] distribution is more appropriately represented by three distinct populations, given the assumption of normally distributed components.
Note that the AIC for a four-component mixture increases to 290, reiterating that more than three populations will overfit the data in hand.
Furthermore, the AIC presumes well-populated Gaussians, for which three components are sufficient to fit the current amount of data.
This analysis does not preclude the possibility of four populations existing when more data are available in the future.
Figure~\ref{fig:km_clustering} depicts the resulting \ri\ and \rii\ classification boundaries, as well as the final three-component Gaussian mixture model.

 \begin{figure}[t]
    \center
    \includegraphics[width=0.9\columnwidth]{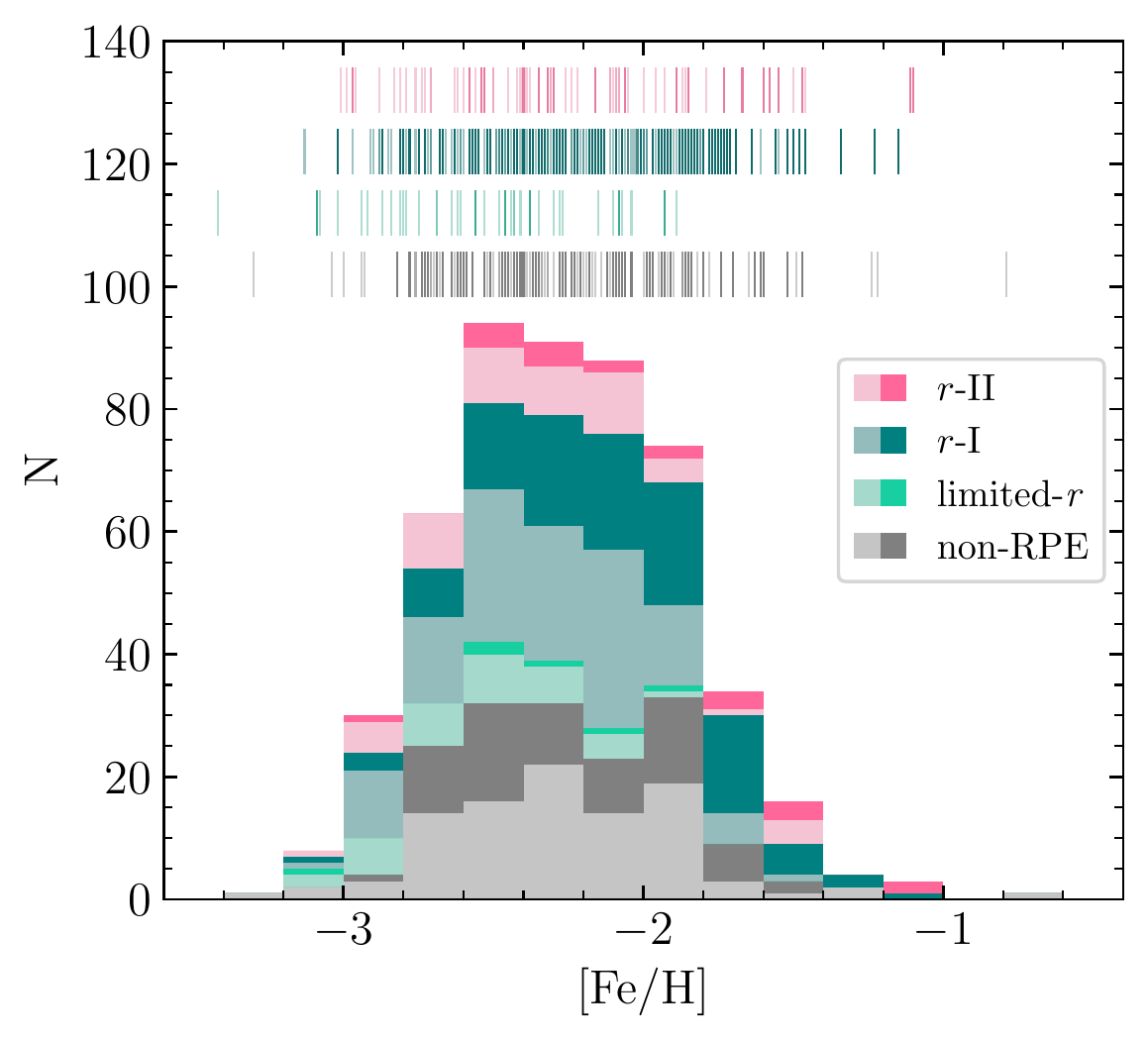}
    \caption{[Fe/H] histograms of \rp-enhanced stars identified by the RPA. Darker colors represent this data release, and lighter colors are all previous RPA work \citep{hansen2018,roederer2018b,sakari2018a,sakari2018b,sakari2019,ezzeddine2020}. At the top, stripe density plots of [Fe/H] for the individual classes are shown.\label{fig:FeH}} 
    \end{figure}

    \begin{figure*}[t]
    \center
    \includegraphics[width=0.9\textwidth]{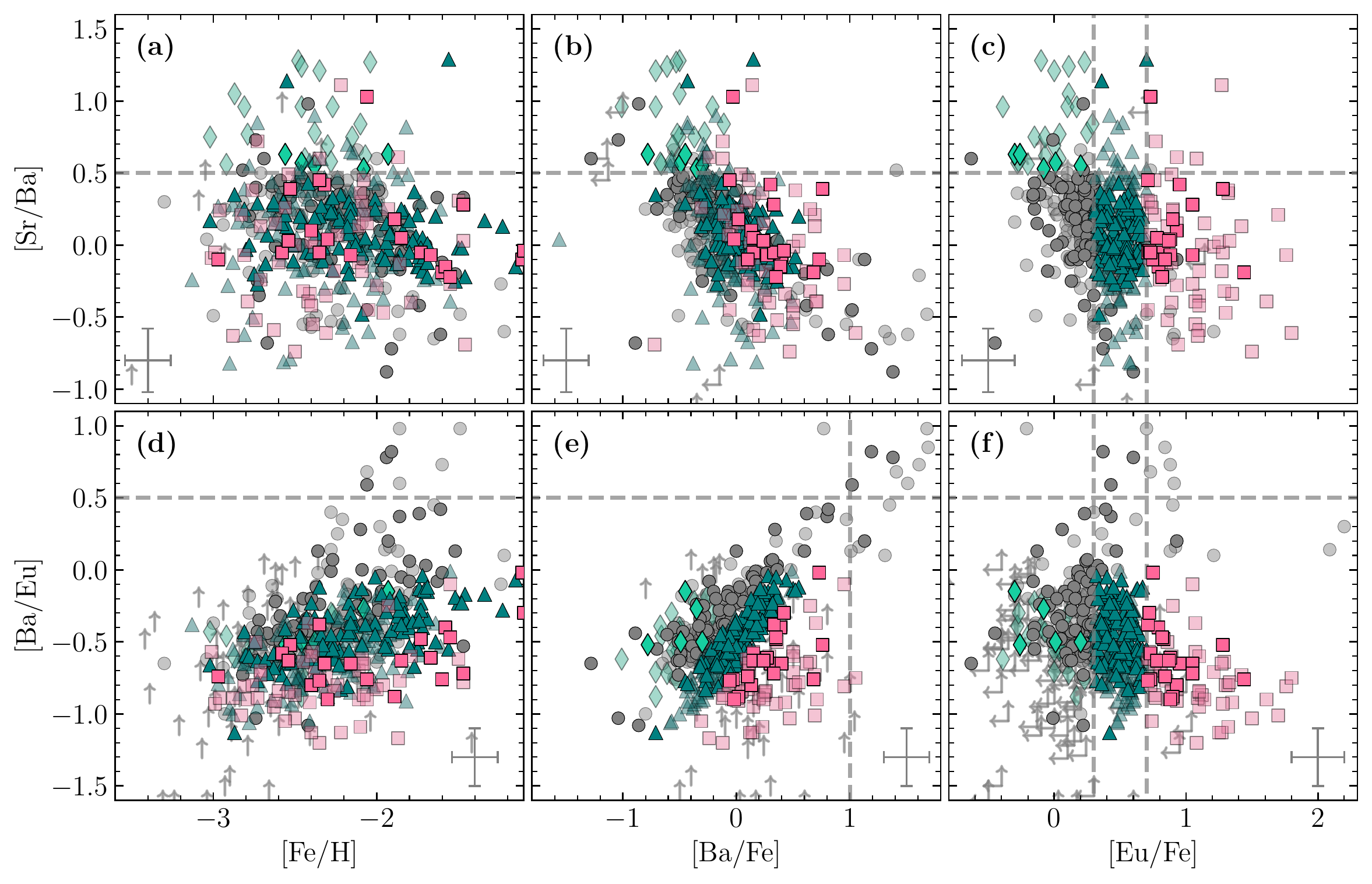}
    \caption{Abundance ratios versus [Fe/H] (panels a and d), [Ba/Fe] (panels b and e), and [Eu/Fe] (panels c and f) for [Sr/Ba] (panels a--c) and [Ba/Eu] (panels d--f) for stars in this sample (dark colors) and previous RPA data releases (light colors). Dashed lines denote the adopted classification cutoffs for \rii, \ri, \limr, and \spr\ signatures (see text for details).\label{fig:SrBa}}
    \end{figure*}

Adopting the split at $[\rm{Eu/Fe}] > +0.7$ to distinguish \rii\ stars from \ri\ stars, the new classifications of \ri\ and \rii\ are now:
	\begin{eqnarray*}
	\mbox{\ri:}\quad& 0.3 < \mathrm{[Eu/Fe]} \leq \mathbf{+0.7},& \quad \mathrm{[Ba/Eu]} < 0\\
    \mbox{\rii:}\quad& \mathrm{[Eu/Fe]} > \mathbf{+0.7},&\quad \mathrm{[Ba/Eu]} < 0.
	\end{eqnarray*}
Note that \citet{roederer2018} also proposed a division at $[\rm{Eu/Fe}]= +0.7$, based on the simple observation that this value effectively excluded most metal-poor stars in the globular cluster and disk populations from the \rii\ class.
This redefinition does not affect the \limr\ class, which are still defined as $[{\rm Eu/Fe}] < +0.3$, $[\rm{Sr/Ba}] > +0.5$, and $[\rm{Sr/Eu}] > 0.0$ as in \citet{frebel2018}.
With this new classification of \ri\ and \rii, we identify a total of 21 new \rii, 111 new \ri, and 7 new \limr\ stars in this data release.
The number of previously identified \ri\ and \rii\ stars (before the RPA was established) changes from 136 and 28 to 99 and 65, respectively, using data in the JINAbase compilation \citep{abohalima2017}.
In the future, as we gather more data about the Milky Way halo, especially at higher metallicities ($[\rm{Fe/H}]\gtrsim -2$), we might consider a metallicity-dependent separation, which may further help distinguish between \rp\ progenitors at different times throughout Galactic history.

Figure \ref{fig:FeH} is a summary of the classification of metal-poor stars based on the RPA data releases to date, using this new definition.
Including this data release, RPA efforts now total 72 \rii, 232 \ri, and 42 \limr\ stars among the 595 targets with snapshot and portrait spectra that have been analyzed to date in \citet{hansen2018,roederer2018b,sakari2018a,sakari2018b,sakari2019} and \citet{ezzeddine2020}.

\subsection{Other Neutron-Capture Signatures}

The [Sr/Ba] and [Ba/Eu] abundance ratios for RPA stars are plotted in Figure~\ref{fig:SrBa} as functions of [Fe/H], [Ba/Fe], and [Eu/Fe].
Stars classified as \limr\ occupy the high-[Sr/Ba], low-[Eu/Fe] end of the scatter (Figure~\ref{fig:SrBa}c).
There are no apparent correlations between the [Sr/Ba] and metallicities for \ri\ and \rii\ stars (Figure~\ref{fig:SrBa}a).
Instead, \ri\ and \rii\ stars are found in roughly equal proportions across a range of low metallicities, implying that the production sites of Sr and Ba are generally uncorrelated in metal-poor stars.
However, we note that all identified \limr\ stars thus far are VMP ($\mathrm{[Fe/H]}\leq -2.0$).
This lack of \limr\ stars at higher metallicities can also be seen in Figure~\ref{fig:EuFe}, in which the spread of [Eu/Fe] abundances dramatically decreases at higher metallicity.
\citet{cote2019} discuss this narrowing in detail, and investigate which \rp\ sites might be responsible for this behavior.
For example, a \limr\ mechanism could have dominated at early times, then became more rare as metallicity increased.

Interestingly, there is a downward trend of high [Sr/Ba] with increasing [Eu/Fe] abundance (Figure~\ref{fig:SrBa}c).
At $\mathrm{[Sr/Ba]} > +0.5$, most stars have somewhat low [Eu/Fe] and are thus classified as \limr\ stars.
Fewer stars have both high [Sr/Ba] and an \ri\ signature, and even fewer stars with high [Sr/Ba] are considered \rii. 
This apparent upper limit could suggest a possible constraint on \limr\ production by prolific main \rp\ sources.
The \rii\ stars with high [Sr/Eu], by definition, show evidence for at least one robust \rp\ source, but also for a potential secondary \limr\ production site, since they exhibit an over-abundance of both Sr and Eu, but a relative under-abundance of Ba.
Our ability to refine and interpret these apparent behaviors will only increase as the size of the RPA sample continues to expand.

We also identify some stars with high [Ba/Eu] and high [Ba/Fe] (Figure~\ref{fig:SrBa}e); the neutron-capture elements in these stars are dominated by \spr\ production.
We identify three new \spr-enhanced stars based on these high Ba ratios, and label them as such in Table~\ref{tab:abund}.
Slightly lower on the [Ba/Eu] scale are stars with a more mixed neutron-capture element signature showing an apparent combination of an \rp\ and \spr\ pattern, notably with $0.0 < \mathrm{[Ba/Eu]}\leq +0.5$ \citep{frebel2018}.
Based on this criterion only, we identify 10 new stars with moderately high [Ba/Eu] abundance ratios.
It is currently unclear how the neutron-capture element abundance pattern in these stars is formed.
For one star it has been identified to be a combination of enrichment by first an \rp\ and then an \spr, earning the label of ``\emph{r}+\emph{s}" (RAVE J094921.8$-$161722; see \citet{gull2018} for details)
For the majority of these stars, this two-component enrichment cannot be invoked to explain their abundance patterns, and it has been speculated to be the signature of the \emph{i}-process \citep{dardelet2015,hampel2016}.
Higher resolution, higher S/N spectroscopic follow-up (``portrait" RPA spectra) on the ten candidate \emph{r}+\emph{s} stars could provide a distinct definition for this new classification of stars, as well as distinguish them from the \emph{r}/\emph{s} and \emph{i}-process classes.

\subsection{Radial-Velocity Variations}

\begin{figure}[t]
    \center
    \includegraphics[width=\columnwidth]{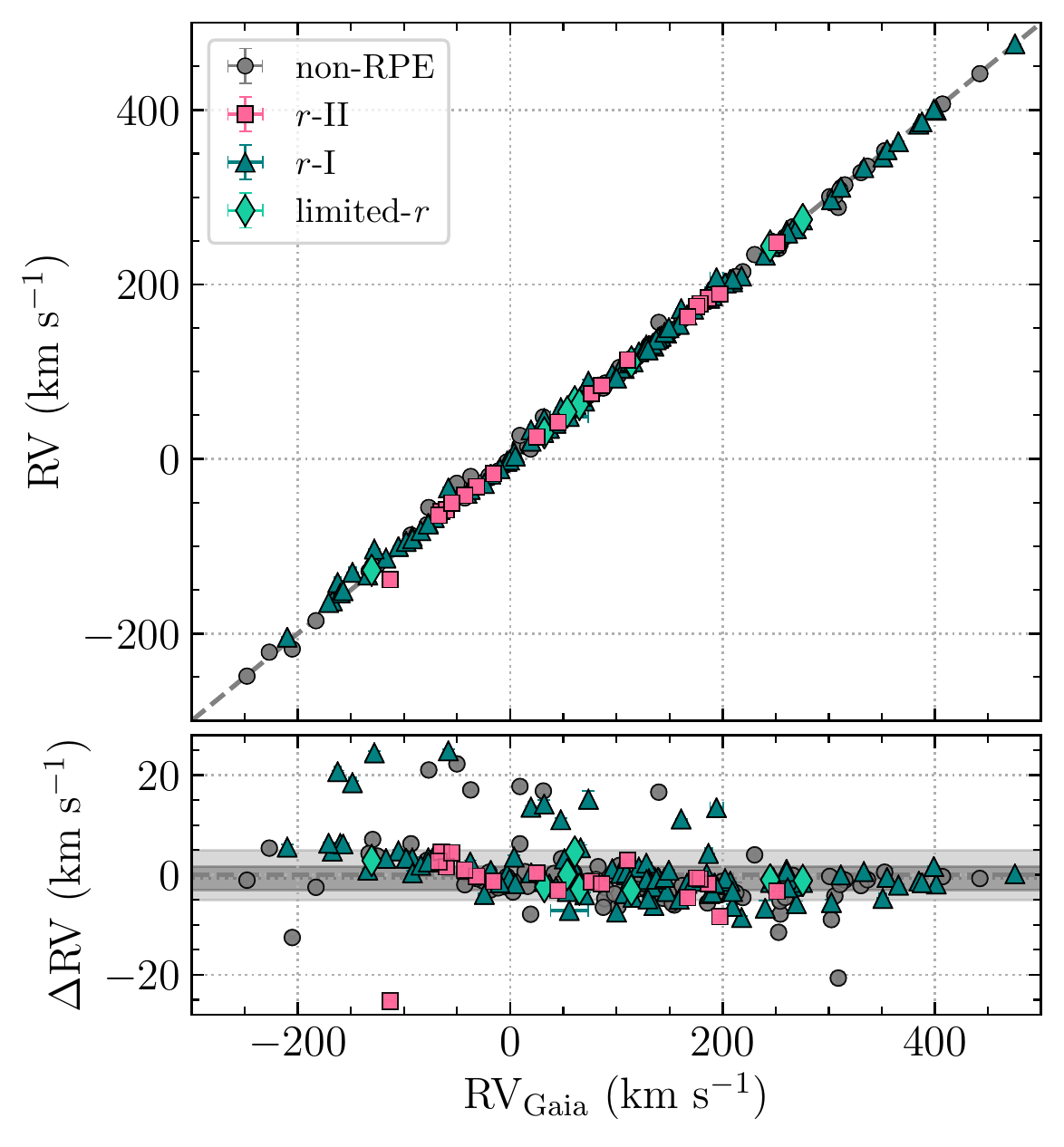}
    \caption{Radial velocities (RVs) reported by Gaia~DR2 for our target stars, compared to RVs measured by this work. Stars with RVs different by more than 5~km\,s$^{-1}$ (outside of the light gray band in the bottom panel) are reported in Table~\ref{tab:possible_binaries}. The dark-gray dotted line and dark-gray band in the lower panel show the average and standard deviation of the residuals for stars within the light-gray band ($-0.64\pm2.35$~km\,s$^{-1}$).\label{fig:rvs_Gaia}}
\end{figure}

The heliocentric RVs measured from our high-resolution spectra are displayed in Figure~\ref{fig:rvs_Gaia}, compared with the Gaia DR2 reported values \citep{gaia2016,gaia2018}.
We find that 47 of our targets (20\%) have a measured RV that differs by more than 5~{km\,s$^{-1}$} from the Gaia DR2 measurement; these stars are listed in Table \ref{tab:possible_binaries}.
In addition to a measurement by {\it Gaia}, many of these stars are found in {\sc Rave} DR5 \citep{kunder2017}, which provides another RV measurement for comparison.
The spread of all RV differences between {\sc Rave} and {\it Gaia} can be fit by two gaussian functions, where the broader gaussian has a standard deviation of 2.6~km\,s$^{-1}$ \citep{steinmetz2020}.
Interestingly, the average RV difference between our high-resolution RV measurements and the {\it Gaia} measurements of non-possible-binary stars is $-0.65\pm 2.35$~km\,s$^{-1}$.
Other spectroscopic surveys also find an negative offset of RVs compared to {\it Gaia} \citep[$-0.3$ for {\sc Rave} and $-$0.2 for APOGEE; see][]{steinmetz2018,steinmetz2020}.
A few other sources of RV estimates for our program stars are available as well and are provided in Table~\ref{tab:possible_binaries}.
Note that most spectroscopic surveys that report an RV (including Gaia DR2 and {\sc Rave}) do not include gravitational redshift or stellar atmosphere corrections, which are expected to be $-0.3\pm 0.2$~km\,s$^{-1}$ for giants \citep{zwitter2018}.
For consistent comparison, such corrections are also omitted from our RV measurements.

Some of the program stars in Table \ref{tab:possible_binaries} suffer from low S/N spectra (e.g., J10540994$-$1347522 and J14165685$+$1215598), increasing the uncertainty in the cross-correlation routine; for completeness, we retain these stars in the list of possible binaries.
Although low S/N effects may lead to a few false positives, the most promising binary candidates are those where the previous literature measurements differ and where the reported {\it Gaia} uncertainty is large ($>$1.0~{km\,s$^{-1}$}).
Note that the {\it Gaia} uncertainty is based on deviations from an average over multiple epochs.
Therefore, a higher uncertainty can indicate possible binarity by reflecting a spread in the individual RV measurements.
On average, the RVs of stars in this RPA data release had eight transits used by Gaia DR2 for their RV measurements.

J04411241$-$6518438 (HD~30229) is a known Pop II field binary with a very low eccentricity and orbital period of about 140 days \citep{pasquini1994}.
J05381700$-$7516207 has several RV measurements, all of which differ from each other outside of their uncertainty.
This star is also an \rii\ star, with an extreme [Eu/Fe] enhancement ($+$1.28), designating it an interesting target for high-resolution follow-up and RV monitoring.
Only one of the possible binaries listed in Table~\ref{tab:possible_binaries} exhibits a high level of carbon enhancement: J03142084$-$1035112, with [C/Fe] = +0.76.
For this star, only upper limits on the Ba and Eu abundances could be determined from our existing spectra.
Detailed follow-up, both with higher-resolution spectroscopy and RV monitoring, can reveal how the observed elemental abundances are affected by mixing and binary interactions, especially for CEMP stars \citep{choplin2019}, to further understand the evolution of the elements.

\startlongtable
\begin{deluxetable*}{lrrDDD}
\tablecaption{Radial Velocities (in {km\,s$^{-1}$}) from Literature and this Data Release for Possible Binaries.\label{tab:possible_binaries}}
\tablehead{
\colhead{2MASS Stellar ID} & \colhead{RV} & \colhead{RV$_{\rm err}$} & \multicolumn2c{RV$_{\rm literature}^\dagger$} & \multicolumn2c{RV$_{\rm err}$}& \multicolumn2c{$\Delta$RV}}
\decimals
\startdata
J00374325$-$1204391  & $-$27.84   &  0.41  & $-$50.14\tablenotemark{a}  &  0.48  & $+$22.30 \\
	& & & $-$48.60\tablenotemark{b}  &  1.74 & $+$20.76 \\
	& & & $-$51.2\tablenotemark{c} & \multicolumn{2}{r}{\nodata} & $+$23.4 \\
J01213447$-$2528002  & $+$33.17   &  0.26  & $+$19.57\tablenotemark{a}  &  2.26  & $+$13.60 \\
	& & & $+$23.62\tablenotemark{b}  &  0.60 & $+$9.55 \\
J01265856$+$0135153  & $-$221.38  &  0.92  & $-$226.79\tablenotemark{a}  &  1.70  & $+$5.41  \\
	& & & $-$211.4\tablenotemark{d} & \multicolumn{2}{r}{\nodata} & $+$10.0 \\
J01311599$-$4016510  & $-$33.31   &  0.38  & $-$58.21\tablenotemark{a}  &  0.28  & $+$24.90 \\
J01371888$-$1729037  & $-$204.31  &  0.55  & $-$209.90\tablenotemark{a}  &  0.77  & $+$5.59  \\
	& & & $-$210.1\tablenotemark{e}  & 0.3 & $+$5.9 \\ 
	& & & $-$209.90\tablenotemark{f} & 0.30 & $+$5.65 \\
J03142084$-$1035112  & $+$241.24  &  0.10  & $+$252.71\tablenotemark{a}  &  4.46  & $-$11.47 \\
	& & & $+$239.3\tablenotemark{d} & \multicolumn{2}{r}{\nodata} & $+$1.9 \\
J03190629$-$0819306  & $+$293.47  &  0.61  & $+$302.39\tablenotemark{a}  &  0.21  & $-$8.92  \\
	& & & $+$304.99\tablenotemark{b} & 2.70 & $-$10.81 \\
J03425812$-$3047217  & $+$296.96  &  0.62  & $+$302.48\tablenotemark{a}  &  0.31  & $-$5.52  \\
	& & & $+$323.8\tablenotemark{d} & \multicolumn{2}{r}{\nodata} & $-$26.8 \\
	& & & $+$302.13\tablenotemark{b} & 0.70 & $-$5.17 \\
J04014897$-$3757533  & $+$156.55  &  0.17  & $+$139.92\tablenotemark{a}  &  0.51  & $+$16.63 \\
	& & & $+$139.77\tablenotemark{b}  &  0.80 & $+$16.78 \\
	& & & $+$139.1\tablenotemark{c} & \multicolumn{2}{r}{\nodata} & $+$17.5 \\
J04315411$-$0632100  & $+$209.37  &  0.27  & $+$217.89\tablenotemark{a}  &  0.43  & $-$8.52  \\
	& & & $+$211.64\tablenotemark{b} & 2.20 & $-$2.27 \\
J04411241$-$6518438  & $+$288.28  &  1.27  & $+$308.93\tablenotemark{a}  &  1.95  & $-$20.65 \\
	& & & $+$292.18\tablenotemark{b} & 1.28 & $-$3.47 \\
J05311779$-$5810048  & $+$129.19  &  0.31  & $+$135.32\tablenotemark{a}  &  1.04  & $-$6.13  \\
	& & & $+$133.38\tablenotemark{b} & 1.56 & $-$4.19 \\
J05381700$-$7516207  & $+$58.73   &  0.34  & $+$47.66\tablenotemark{a}  &  1.05  & $+$11.07 \\
	& & & $+$52.91\tablenotemark{b} & 0.68 & $+$12.15 \\
	& & & $+$43.0\tablenotemark{c} & \multicolumn{2}{r}{\nodata} & $+$22.1 \\
J05383296$-$5904280  & $+$189.02  &  0.37  & $+$197.35\tablenotemark{a}  &  0.41  & $-$8.33  \\
	& & & $+$196.65\tablenotemark{b} & 0.54 & $-$7.63 \\
J06014757$-$5951510  & $+$246.53  &  0.22  & $+$254.34\tablenotemark{a}  &  0.31  & $-$7.81  \\
	& & & $+$253.64\tablenotemark{b} & 0.56 & $-$7.11 \\
J06420823$-$5116448  & $+$15.43   &  0.27  & $+$9.16\tablenotemark{a}  &  0.80  & $+$6.27  \\
	& & & $+$8.57\tablenotemark{b} & 0.87 & $+6.86 \\
J07265723$-$5647500  & $+$71.75   &  0.62  & $+$66.31\tablenotemark{a}  &  3.33  & $+$5.44  \\
	& & & $+$63.29\tablenotemark{b} & 1.11 & $+$9.75 \\
J09255655$-$3450373  & $+$203.06  &  0.39  & $+$209.35\tablenotemark{a}  &  0.27  & $-$6.29  \\
	& & & $+$210.11\tablenotemark{b} & 0.59 & $-$7.05 \\
J10025125$-$4331098  & $+$233.46  &  1.53  & $+$240.13\tablenotemark{a}  &  1.93  & $-$6.67  \\
J10251539$-$3554026  & $+$248.87  &  0.45  & $+$254.10\tablenotemark{a}  &  0.59  & $-$5.23  \\
	& & & $+$254.68\tablenotemark{b} & 3.62 & $-$5.81 \\
J10302845$-$7543299  & $+$263.78  &  0.54  & $+$269.53\tablenotemark{a}  &  0.31  & $-$5.75  \\
	& & & $+$270.29\tablenotemark{b}  & 0.86 & $-$6.51 \\
J10540994$-$1347522  & $+$180.23  &  0.77  & $+$185.81\tablenotemark{a}  &  1.76  & $-$5.58  \\
	& & & $+$188.7\tablenotemark{d} & \multicolumn{2}{r}{\nodata} & $-$8.5 \\
J11404726$-$0833030  & $+$172.22  &  0.10  & $+$161.00\tablenotemark{a}  &  1.63  & $+$11.22 \\
	& & & $+$160.33\tablenotemark{b} & 0.84 & $+$11.89 \\
J14165685$+$1215598  & $-$87.15   &  0.82  & $-$93.43\tablenotemark{a}  &  2.10  & $+$6.28  \\
J15141994$-$4359554  & $+$148.50  &  0.97  & $+$154.50\tablenotemark{a}  &  0.98  & $-$6.00  \\
J15360493$+$0247300  & $-$20.03   &  0.23  & $-$37.11\tablenotemark{a}  &  2.56  & $+$17.08 \\
J19050116$-$1949280  & $+$95.31   &  0.35  & $+$101.60\tablenotemark{a}  &  0.57  & $-$6.29  \\
	& & & $+$99.51\tablenotemark{b} & 1.38 & $-$4.20 \\
J19175585$-$5440147  & $+$48.08   &  0.54  & $+$31.22\tablenotemark{a}  &  0.68  & $+$16.86 \\
	& & & $+$25.51\tablenotemark{b} & 3.94 & $+$22.47 \\
J19445483$-$4039459  & $+$92.74   &  0.30  & $+$100.17\tablenotemark{a}  &  0.53  & $-$7.43  \\
	& & & $+$99.52\tablenotemark{b} & 0.58 & $-$6.78 \\
J19451414$-$1729269  & $+$46.23   &  0.80  & $+$32.00\tablenotemark{a}  &  1.66  & $+$14.23 \\
	& & & $+$35.45\tablenotemark{b} & 0.54 & $+$10.78 \\
	& & & $+$30.4\tablenotemark{g} & \multicolumn{2}{r}{\nodata} & $+$15.8 \\
	& & & $+$30.60\tablenotemark{f} & 0.20 & $+$15.63 \\
J20194310$-$3158163  & $-$130.00  &  0.30  & $-$148.51\tablenotemark{a}  &  3.34  & $+$18.51 \\
	& & & $-$153.43\tablenotemark{b} & 0.56 & $+$23.43 \\
J20233743$-$1659533  & $-$141.72  &  0.23  & $-$162.43\tablenotemark{a}  &  3.05  & $+$20.71 \\
	& & & $-$157.44\tablenotemark{b} & 1.34 & $+$15.72 \\
J20504869$-$3355289  & $-$153.72  &  0.22  & $-$160.02\tablenotemark{a}  &  0.39  & $+$6.30  \\
	& & & $-$158.76\tablenotemark{b} & 0.56 & $+$5.04 \\
J20554594$-$3155159  & $-$151.07  &  0.19  & $-$157.29\tablenotemark{a}  &  0.25  & $+$6.22  \\
	& & & $-$155.06\tablenotemark{b} & 0.65 & $+$3.99 \\
J21055865$-$4919336  & $+$207.72  &  0.19  & $+$194.24\tablenotemark{a}  &  6.10  & $+$13.48 \\
	& & & $+$169.90\tablenotemark{b} & 1.71 & $+$37.82 \\
J21080151$-$6555366  & $+$81.27   &  0.27  & $+$87.72\tablenotemark{a}  &  0.39  & $-$6.45  \\
	& & & $+$88.44\tablenotemark{b} & 0.38 & $-$7.17 \\
J21103411$-$6331354  & $-$122.35  &  0.23  & $-$129.48\tablenotemark{a}  &  0.44  & $+$7.13  \\
	& & & $-$130.56\tablenotemark{b} & 1.54 & $+$8.21 \\
J21314253$-$1459110  & $+$11.39   &  0.30  & $+$19.24\tablenotemark{a}  &  2.87  & $-$7.85  \\
	& & & $+$17.79\tablenotemark{b} & 0.91 & $-$6.40 \\
J22125424$-$0235414  & $-$103.40  &  0.40  & $-$127.91\tablenotemark{a}  &  0.95  & $+$24.51 \\
	& & & $-$127.05\tablenotemark{b} & 1.29 & $+$23.65 \\
	& & & $-$145.8\tablenotemark{i} & \multicolumn{2}{r}{\nodata} & $+$42.4 \\
J22161170$-$5319492  & $+$88.90   &  1.71  & $+$73.70\tablenotemark{a}  &  0.44  & $+$15.20 \\
	& & & $+$73.39\tablenotemark{b} & 1.69 & $+$15.51 \\
J22223324$-$1314488  & $+$26.96   &  0.25  & $+$9.19 \tablenotemark{a}  &  0.55  & $+$17.77 \\
	& & & $+$11.07\tablenotemark{b} & 0.73 & $+$15.89 \\
J22233596$-$5301145  & $+$146.88  &  0.26  & $+$152.56\tablenotemark{a}  &  0.29  & $-$5.68  \\
	& & & $+$153.6\tablenotemark{h} & 1.4 & $-$6.7 \\
J22372037$-$4741375  & $-$138.25  &  0.26  & $-$112.98\tablenotemark{a}  &  4.45  & $-$25.27 \\
	& & & $-$107.5\tablenotemark{d} & \multicolumn{2}{r}{\nodata} & $-$30.8 \\
J22585069$-$3923437  & $-$55.60   &  0.40  & $-$76.68\tablenotemark{a}  &  1.63  & $+$21.08 \\
J23425814$-$4327352  & $+$48.53   &  0.22  & $+$55.66\tablenotemark{a}  &  17.69  & $-$7.13  \\
J23490902$-$2447176  & $-$164.58  &  0.22  & $-$170.94\tablenotemark{a}  &  0.68  & $+$6.36  \\
	& & & $-$167.46\tablenotemark{b} & 0.90 & $+$2.88 \\ 
J23552837$+$0421179  & $-$217.70  &  0.41  & $-$205.18\tablenotemark{a}  &  3.38  & $-$12.52 \\
\enddata
\tablenotetext{\dagger}{Sources are defined as follows: \emph{a}: Gaia DR2 \cite{gaia2018}; \emph{b}: RAVE DR5 \citet{kunder2017}; \emph{c}: \citet{ruchti2011}; \emph{d}: \citet{beers2017}; \emph{e}: \citet{ishigaki2012}; \emph{f}: \citet{gontcharov2006}; \emph{g}: \citet{roederer2014a}; \emph{h}: RAVE DR3 \citet{siebert2011}; and \emph{i}: \citet{schlaufman2014}.}
\end{deluxetable*}

\section{Summary and Discussion}
\label{discussion}

This data set constitutes the fourth data release of the RPA search for \rpe\ stars, culminating in a current total of 595 metal-poor stars with Phase II (snapshot) spectroscopy in the total published sample \citep{hansen2018,roederer2018b,sakari2018a,sakari2018b,sakari2019,ezzeddine2020}.
Another $\sim$1000 snapshot spectra of candidates have already been taken with a number of telescopes in the Northern and Southern Hemispheres and will be released in due course.

Quantified chemical identifications provide clues as to the formation history of the Milky Way, since stars with similar metallicities and levels of \rp\ enrichment have also been found to be dynamically linked in small associations \citep[see, e.g.,][]{roederer2018,yuan2019}.
Current and future RPA efforts will help to refine the mapping of \rp-enhanced stars into their parent dynamical groups, so that we may learn more about the natal environment in which the \rp\ occurred in each of these now-disrupted systems.
By identifying entire systems of \rpe\ stars that likely shared a common birthplace and star-formation history, we can test the dilution hypothesis of nucleosynthetic events, i.e., whether the heavy-element material in \ri\ and \rii\ stars came from similar sources, but the \ri\ stars have simply been diluted by larger masses of baryons in their natal mini-halos, leading to smaller enhancements.
\citet{tarumi2020} suggest other alternatives to account for the different levels of \rp\ enhancements in the UFDs Ret~II and Tuc-III (and by extension to the \ri\ and \rii\ stars in the halo field) based on the locations of their progenitor NSMs.

Future data releases by the RPA will continue to increase the number of stars with identified \rp\ signatures, and perhaps reveal new ones for investigations of the various proposed nucleosynthetic sites.
Fresh investigations of actinide production, for example, are being used to distinguish between specific \rp\ sites and the conditions that produce these heavy elements \citep{eichler2019,holmbeck2019b}.
Furthermore, the identification of dynamical groups that include \rpe\ stars are useful to constrain theoretical models of \rp\ production; see, e.g., \citet{holmbeck2019a} and Gudin et al.\ (in prep.).

\acknowledgments{
This publication is based upon work supported in part by the U.S.\ National Science Foundation (NSF) under grant AST-1714873. E.M.H, T.C.B, V.M.P, K.C.R., D.D.W., I.U.R, and A.F.\ acknowledge partial support from grant PHY~14-30152 (Physics Frontier Center/JINA-CEE), awarded by the NSF.
I.U.R.\ acknowledges support from NSF grants 1613536 and 1815403.
A.F.\ acknowledges support from NSF CAREER grant AST-1255160.
E.K.G.\ acknowledges funding by the Deutsche Forschungsgemeinschaft (DFG, German Research Foundation), Project-ID 138713538 -- SFB 881 (``The Milky Way System", subprojects A03 and A05).
Funding for {\sc Rave} has been provided by: the Australian Astronomical Observatory; the Leibniz-Institut fuer Astrophysik Potsdam (AIP); the Australian National University; the Australian Research Council; the French National Research Agency; the German Research Foundation (SPP 1177 and SFB 881); the European Research Council (ERC-StG 240271 Galactica); the Istituto Nazionale di Astrofisica at Padova; The Johns Hopkins University; the NSF (AST-0908326); the W.\ M.\ Keck foundation; the Macquarie University; the Netherlands Research School for Astronomy; the Natural Sciences and Engineering Research Council of Canada; the Slovenian Research Agency; the Swiss National Science Foundation; the Science \& Technology Facilities Council of the UK; Opticon; Strasbourg Observatory; and the Universities of Groningen, Heidelberg and Sydney. The {\sc Rave} web site is at \url{https://www.rave-survey.org}.
This work has made use of data from the European Space Agency (ESA) mission {\it Gaia} (\url{https://www.cosmos.esa.int/gaia}), processed by the {\it Gaia} Data Processing and Analysis Consortium (DPAC, \url{https://www.cosmos.esa.int/web/gaia/dpac/consortium}). Funding for the DPAC has been provided by national institutions, in particular the institutions participating in the {\it Gaia} Multilateral Agreement.
This research has made use of NASA's Astrophysics Data System Bibliographic Services.
}

\facilities{du Pont 2.5m telescope}
\software{
\texttt{matplotlib} \citep{hunter2007}, \texttt{CarPy} \citep{kelson1998,kelson2000,kelson2003},  \texttt{IRAF} \citep{tody1986,tody1993}, \texttt{MOOG} \citep{sneden1973}, \texttt{linemake} (\url{https://github.com/vmplacco/linemake}), \texttt{ATLAS9} \citep{castelli2003}
}

\bibliography{bibliography.bib}

\end{document}